\documentclass[esd, manuscript]{copernicus}

\nolinenumbers
\usepackage{todonotes}

\usepackage{xcolor}

\begin{document}

\title{Earth system modeling with endogenous and dynamic human societies: the copan:CORE open World-Earth modeling framework}

\Author[1,2,*]{Jonathan F.}{Donges}
\Author[1,*]{Jobst}{Heitzig}
\Author[1,3]{Wolfram}{Barfuss}
\Author[1]{Marc}{Wiedermann}
\Author[1,4]{Johannes A.}{Kassel}
\Author[3]{Tim}{Kittel}
\Author[1,3]{Jakob J.}{Kolb}
\Author[1,3]{Till}{Kolster}
\Author[1,3]{Finn}{M\"uller-Hansen}
\Author[1]{Ilona M.}{Otto}
\Author[1,5]{Kilian B.}{Zimmerer}
\Author[1,6,7]{Wolfgang}{Lucht}

\affil[1]{Earth System Analysis and Complexity Science, Potsdam Institute for Climate Impact Research, Member of the Leibniz Association, Telegrafenberg A31, D-14473 Potsdam, Germany}
\affil[2]{Stockholm Resilience Centre, Stockholm University, Kr\"aftriket 2B, 114 19 Stockholm, Sweden}
\affil[3]{Department of Physics, Humboldt University, Newtonstr. 15, D-12489 Berlin, Germany}
\affil[4]{Department of Physics, University of Göttingen, Friedrich-Hund-Platz 1, D-37077 Göttingen, Germany}
\affil[5]{Department of Physics and Astronomy, University of Heidelberg, Im Neuenheimer Feld 226, D-69120 Heidelberg, Germany}
\affil[6]{Department of Geography, Humboldt University, Unter den Linden 6, D-10099 Berlin, Germany}
\affil[7]{Integrative Research Institute on Transformations of Human-Environment Systems, Humboldt University, Unter den Linden 6, D-10099 Berlin, Germany}
\affil[*]{The first two authors share the lead authorship.}


\runningtitle{The copan:CORE open World-Earth modeling framework}

\runningauthor{Heitzig, Donges et al.}  

\correspondence{Jonathan F. Donges (donges@pik-potsdam.de), Jobst Heitzig (heitzig@pik-potsdam.de)}

\received{}
\pubdiscuss{} 
\revised{}
\accepted{}
\published{}

\firstpage{1}

\maketitle

\begin{abstract}

Analysis of Earth system dynamics in the Anthropocene requires to explicitly take into account the increasing magnitude of processes operating in human societies, their cultures, economies and technosphere and their growing feedback entanglement with those in the physical, chemical and biological systems of the planet. However, current state-of-the-art Earth System Models do not represent dynamic human societies and their feedback interactions with the biogeophysical Earth system and macroeconomic Integrated Assessment Models typically do so only with limited scope. 
This paper (i) proposes design principles for constructing World-Earth Models (WEM) for Earth system analysis of the Anthropocene,
i.e., models of social (World) - ecological (Earth) co-evolution on up to planetary scales, and (ii) presents the copan:CORE open
simulation modeling framework 
for developing, composing and analyzing such WEMs based on the proposed principles. 
The framework provides a modular structure to flexibly construct and study WEMs. These can contain biophysical (e.g. carbon cycle dynamics), socio-metabolic/economic (e.g. economic growth or energy system changes) and socio-cultural processes (e.g. voting on climate policies or changing social norms) and their feedback interactions, and are based on elementary entity types, e.g., grid cells and social systems. 
Thereby, copan:CORE enables the epistemic flexibility needed for contributions towards Earth system analysis of the Anthropocene given the large diversity of competing theories and methodologies used for describing socio-metabolic/economic and socio-cultural processes in the Earth system by various fields and schools of thought.
To illustrate the capabilities of the framework, we present an exemplary and highly stylized WEM implemented in copan:CORE that illustrates how endogenizing socio-cultural processes and feedbacks such as voting on climate policies based on socially learned environmental awareness could fundamentally change macroscopic model outcomes.
\end{abstract}


%
%
\section{Introduction}
\label{sec:intro}

In the Anthropocene, Earth system dynamics is equally governed 
by two kinds of internal processes: 
those operating in the physical, chemical, and biological systems of the planet and those occurring in its human societies, 
their cultures and economies~\citep{Schellnhuber1998,Schellnhuber1999,Crutzen2002geology,Steffen2017trajectories}.
The history of global change is the history of the increasing planetary-scale entanglement and strengthening of feedbacks between these two domains \citep{lenton2011revolutions}. Therefore, Earth system analysis of the Anthropocene requires to close the loop by integrating the dynamics of complex human societies into integrated \emph{whole} Earth system models~\citep{Verburg2016,Donges2017a,Donges2017b}.
Such models need to capture the co-evolving dynamics of the social (the \emph{World} of human societies) 
and natural (the biogeophysical \emph{Earth}) spheres of the Earth system on up to global scales and are referred to as World-Earth models (WEMs) in this article. In pursuing this interdisciplinary integration effort, World-Earth modeling can benefit from and build upon the work done in fields such as social-ecological systems~\citep{berkes2000linking,folke2006} and coupled human and natural systems~\citep{liu2007complexity} research or land-use~\citep{arneth2014global} and socio-hydrological modeling~\citep{dibaldassarre2017}. However, it emphasizes more the study of planetary scale interactions between human societies and parts of the Earth's climate system such as atmosphere, ocean and the biosphere, instead of more local and regional scale interactions with natural resources that these fields have typically focused on in the past~\citep{Donges2017c}.

The contribution of this paper is twofold:  
First, following a more detailed motivation (Sect.\,\ref{sec:motivation}), general theoretical considerations and design principles for a novel class of integrated WEMs are discussed (Sect.\,\ref{sec:blueprinting}) and WEMs are discussed in the context of existing global modelling approaches (Sect.\,\ref{sec:wem_comparison}).
Second, after a short overview of the copan:CORE open World-Earth modeling framework (Sect.\,\ref{sec:model_description}), an exemplary full-loop WEM is presented and studied (Sect.\,\ref{sec:application}), showing the relevance of internalizing socio-cultural processes. Finally, Sect.\,\ref{sec:conclusions} concludes the paper.

\subsection{State of the art and research gaps in Earth system analysis}
\label{sec:motivation}


Computer simulation models are pivotal tools for gaining scientific understanding 
and providing policy advice for addressing global change challenges 
such as anthropogenic climate change or rapid degradation of biosphere integrity and their interactions~\citep{Rockstrom2009safe,Steffen2015planetary}.  
At present, two large modeling enterprises considering the larger Earth system in the Anthropocene are mature~\citep{vanVuuren2016}:
(i)~Biophysical ``Earth system models'' (ESMs) derived from and built around a core of atmosphere-ocean general circulation models 
that are evaluated using storyline-based socioeconomic scenarios to study anthropogenic climate change and its impacts on human societies (e.g., representative concentration pathways, RCPs)~\citep{IPCC2013WG1}.
(ii)~Socio-economic Integrated Assessment Models (IAMs) are operated using storyline-based socio-economic baseline scenarios (e.g., shared socio-economic pathways, SSPs, \citet{IPCC2014WG3}) and evaluate technology and policy options for mitigation and adaption leading to different emission pathways.
There is a growing number of intersections, couplings and exchanges between 
the biophysical and socio-economic components of
these two model classes for increasing their consistency~\citep{vanVuuren2012,foley2016,dermody2017,Robinson2018}.

\begin{figure}[t]
\includegraphics[width=.75 \textwidth]{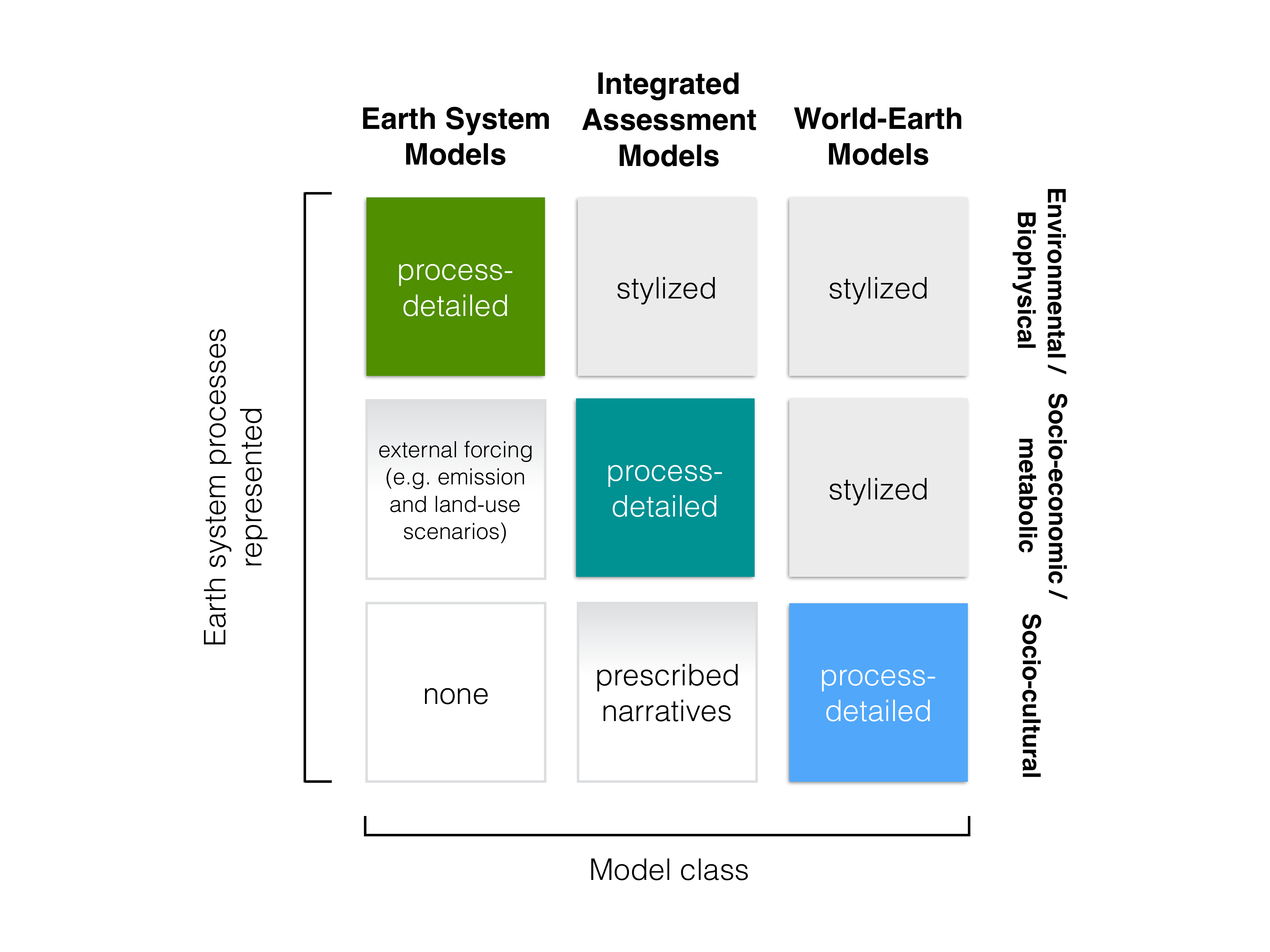}
\caption{%
	\textbf{World-Earth models (WEMs) in the space of model classes used for scientific analysis of global change.} 
    It is shown to what degree current Earth system models, 
    integrated assessment models and WEMs 
    cover environmental/biophysical, socio-economic/metabolic, and socio-cultural processes, respectively.
    The term ``process-detailed" indicates the types of Earth system processes that the different model classes typically focus on representing. However, also in these core areas the level of detail may range from very stylized to complex and highly structured.
}
\label{fig:matrix}
\end{figure}

However, the existing scientific assessment models of global change 
include only to a limited degree -- if at all -- dynamic representations of the socio-cultural dimensions of human societies (Fig.~\ref{fig:matrix}), i.e. the diverse political and economic actors, the factors influencing their decisions and behavior, their interdependencies constituting social network structures and institutions~\citep{Verburg2016,Donges2017a,Donges2017b} as well as the broader technosphere they created~\citep{haff2012technology,haff2014humans}.
In IAMs, these socio-cultural dimensions are partly represented by different socio-economic scenarios 
(e.g., SSPs), providing the bases for different emission pathways.
These are in turn used in ESMs as external forcing, constraints and boundary conditions to the modeled Earth system dynamics.
However, a dynamic representation would be needed to explore how changes in the global environment influence these socio-cultural factors and vice versa.

There are large differences in beliefs, norms, economic interests, and political ideologies of various social groups, 
and their metabolic profiles, which are related to their access and use of energy and resources~\citep{fischerkowalski1997,otto2019shift,lenton2016revolutions,lenton2018gaia}. Historical examples show that these differences might lead to rapid social changes, revolutions and sometimes also devastating conflicts, wars and collapse \citep{betts2017conflict,cumming2017unifying}.
In other cases, the inability to establish effective social institutions 
controlling resource access might lead to unsustainable resource use and resource degradation~\citep[see the discussion around the tragedy of the commons,][]{Ostrom1990,jager2000behaviour,janssen2002complexity}.
Climate change is a paradigmatic example of a global commons 
that needs global institutional arrangements for the usage of the atmosphere as a deposit for greenhouse gas emissions 
if substantial environmental and social damages are to be avoided in the future~\citep{edenhofer2015atmosphere,Schellnhuber2016challenge,otto2017social}.

In order to explore the risks, dangers and opportunities for sustainable development,
it is important to understand how biophysical, socio-economic and socio-cultural processes influence each other~\citep{Donges2017c}, 
how institutional and other social processes function, 
and which tipping elements can emerge out of the interrelations of the 
subsystems~\citep{lenton2008tipping,kriegler2009,cai2016risk,kopp2016tipping}.
To address these questions, the interactions of social systems and the natural Earth system can be regarded as part of a planetary social-ecological system (SES) or World-Earth system, extending the notion of SES beyond its common usage to describe systems on local scales~\citep{berkes2000linking,folke2006}.
This dynamical systems perspective allows to explore
under which preconditions the maintenance of planetary boundaries~\citep{Rockstrom2009safe,Steffen2015planetary}, 
i.e., a Holocene-like state of the natural Earth System, 
can be reconciled with human development
to produce an ethically defensible trajectory of the whole Earth system (i.e., sustainable development)~\citep{Raworth2012,Steffen2017trajectories}.

\subsection{World-Earth modeling: contributions towards Earth system analysis of the Anthropocene}
\label{sec:blueprinting}

To this end, the case has been made that substantial efforts are required to advance the development of integrated World-Earth system models~\citep{Verburg2016,Donges2017a,Donges2017b}. 
The need for developing such next generation social-ecological models has been recognized in several subdisciplines of global change science dealing with socio-hydrology~\citep{dibaldassarre2017,keys2017}, land-use dynamics~\citep{arneth2014global,Robinson2018}, and the globalized food-water-climate nexus~\citep{dermody2017}.
While in recent years there has been some progress in developing stylized models that combine socio-cultural with economic and natural dynamics (e.g. \citet{janssen1998battle,Kellie2011,garrett2014long,Motesharrei2014,Wiedermann2015,heck2016collateral,barfuss2017sustainable,Nitzbon2017}),
more advanced and process-detailed WEMs are not yet available for studying 
the deeper past and the longer-term Anthropocene future of this coupled system.
The research program investigating the dynamics and resilience of the World-Earth system in the Anthropocene 
can benefit from recent advances in the theory and modeling of complex adaptive systems~\citep{farmer2015third,Verburg2016,Donges2017a,Donges2017b}.
When going beyond stylized modelling, a key challenge for World-Earth modeling is the need to take into account the agency of heterogeneous social actors 
and global-scale adaptive networks carrying and connecting 
social, economic and ecological processes 
that shape social-ecological co-evolution.

A number of new developments make it attractive to re-visit the challenge of building such WEMs now.
Due to the huge progress in computing, comprehensive Earth system modeling is advancing fast.
And with the ubiquity of computers and digital communication for simulation and data acquisition in daily life~\citep{otto2015socio}, 
efforts to model complex social systems are increased and become more concrete. 
Recent advances for example in complex systems theory, computational social sciences, social simulation and social-ecological systems modeling~\citep{farmer2009economy,farmer2015third,helbing2012futurict,Muller-Hansen2017} 
make it feasible to include some important macroscopic dynamics of human societies regarding among others 
the formation of institutions, values, and preferences, and various processes of decision-making 
into a model of the whole Earth system, i.e., the physical Earth including its socially organised and mentally reflexive humans.
Furthermore, new methodological approaches are developing fast 
that allow representing crucial aspects of social systems, 
such as adaptive complex networks~\citep{gross2008adaptive,snijders2010introduction}.
Finally, initiatives such as {\em Future Earth}~\citep{futureearth2014} and the {\em Earth League}~(\citet{rockstrom2014climate}, www.the-earth-league.org)
provide 
a basis 
for inter- and trans-disciplinary research that could support such an ambitious modeling program.


\subsubsection{Research questions for World-Earth modeling}

We envision World-Earth modeling to be complementary to existing simulation approaches for the analysis of global change. WEMs are not needed where the focus is on the study of the biophysical and climatic implications of certain prescribed socio-economic development pathways (e.g. in terms of emission and land-use scenarios), since this is the domain of Earth System Models as used in the World Climate Research Programme's Coupled Model Intercomparison Project (CMIP)~\citep{eyring2016overview} that provides input to the Intergovernmental Panel on Climate Change (IPCC) reports. Similarly, WEMs are not the tool of choice if the interest is in the normative macro-economic projection of optimal socio-economic development and policy pathways internalizing certain aspects of climate dynamics, e.g. the analysis of first or second best climate change mitigation pathways, since this is the domain of state-of-the-art Integrated Assessment Models.

In turn, WEMs as envisioned by us here are needed when the research questions at hand require the explicit and internalized representation of socio-cultural processes and their feedback interactions with biophysical and socio-economic dynamics in the Earth system. In the following, we give examples for research questions of this type that could be studied with WEMs in the future, as they have been already elaborated in more detail by, e.g. \citet{Verburg2016} and \citet{Donges2017a,Donges2017b}:

\begin{enumerate}

\item What are the socio-cultural, -economic and environmental preconditions for sustainable development 
towards and within a ``safe and just'' operating space for humankind~\citep{barfuss2018optimization,o2018good}, 
i.e., for a trajectory of the Earth system that eventually neither violates precautionary planetary boundaries~\citep{Rockstrom2009safe,Steffen2015planetary}
nor acceptable social foundations~\citep{Raworth2012}?


\item A more specific example of the previous questions is: How can major socio-economic transitions towards a decarbonized social metabolism, such as a transformation of the food and agriculture system towards a sustainable, reduced-meat diet that is in line with recent recommendations by the EAT-Lancet Commission on healthy diets \citep{willett2019food}, be brought about in view of the strong socio-cultural drivers of current food-related and agricultural practises and the reality of the political economy in major food-producing countries? And how would their progress be influenced by realized or anticipated tipping of climatic tipping elements like the Indian monsoon system?

\item Under which conditions can cascading interactions between climatic (e.g., continental ice sheets or major biomes such as the Amazon rain forest) 
and potential social tipping elements (e.g., in attitudes towards ongoing or anticipated climate change or eco-migration) be triggered and how can they be governed~\citep{schellnhuber2016right,Steffen2017trajectories}? What are implications for biophysical and social-ecological dimensions of Earth system resilience in the Anthropocene~\citep{Donges2017a}?

\item How do multilevel coalition formation processes (like the one modeled in \cite{heitzig2018bottom} assuming a static climate) interact with Earth system dynamics via changes in regional damage functions, mitigation costs, and realized or anticipated distributions of extreme events that drive changes in public opinions which in turn influence the ratification of international treaties and the implementation of domestic climate policies?




\item How do certain social innovations including technology, policies or behavioral practises diffuse in heterogeneous agent networks that could have global-scale impacts on planetary-boundary dimensions (e.g. \citet{farmer2019sensitive,tabara2018positive})? Which factors such as network structure, information access as well as information feedback and update time affect the innovation uptake? What are the impacts of a certain social innovation uptake on different agent groups (e.g. on agents with different economic, social, or cultural endowment)? \citep{hewitt2019social} 



\end{enumerate}

\subsubsection{Design principles for World-Earth models}

To address research questions of the kinds suggested by the examples given above, we suggest that the development of WEMs of the type discussed in this paper could be guided by aiming for the following properties:

\begin{enumerate}

    \item \textbf{Explicit representation of social dynamics.}
    Societal processes should be represented in an explicit, dynamic fashion in order to do justice to the dominant role of human societies in the Anthropocene.
    (In contrast, social process occur typically non-dynamically in ESMs as fixed socio-economic pathways; and in IAMs as inter-temporal optimization problem.)
    %
    Such social processes such as social learning may be included in models via comparably simple equation-based descriptions (e.g.~\citet{Donges2017c}).
    Yet, more detailed WEMs should also allow also for representations of the dynamics of 
    the diverse agents and the complex social structure connecting them that constitute human societies, 
    using the tools of agent-based and adaptive network modeling~\citep{farmer2009economy,farmer2015third,Muller-Hansen2017}. 
    The social sphere is networked on multiple layers and regarding multiple phenomena (knowledge, trade, institutions, preferences etc.) and that increasing density of such interacting network structures is one of the defining characteristics of the Anthropocene~\citep{Steffen2007anthropocene,gaffney2017anthropocene}.
    While there is a rich literature on modeling various aspects of socio-cultural dynamics (e.g. \citet{castellano2009statistical,snijders2010introduction,Muller-Hansen2017,schluter2017framework}), this work so far remains mostly disconnected from Earth system modeling. 
    Accordingly, more detailed WEMs should be able to describe decision processes of representative samples of 
    individual humans, social groups or classes, and collective agents such as firms, households or governments. 
    This includes the representation of diverse objectives, constraints, and decision rules, differentiating for example by the agent's social class and function 
    and taking the actual and perceived decision options of different agent types into account. 
    
    \item \textbf{Feedbacks and co-evolutionary dynamics} WEMs should incorporate as dynamic processes 
    the feedbacks of collective social processes on biogeophysical Earth system components and vice versa.
    The rationale behind this principle is that the strengthening of such feedbacks, e.g. the feedback loop consisting of anthropogenic greenhouse gas emissions driving climate change acting back on human societies through increasingly frequent extreme events, is one of the key characteristics of the Anthropocene. Moreover, the ability to simulate feedbacks is central to a social-ecological and complex adaptive systems approach to Earth system analysis.
    Capturing these feedbacks enables them to produce paths in co-evolution space~\citep{Schellnhuber1998,Schellnhuber1999} 
    through time-forward integration of all entities and networks 
    allowing for deterministic and stochastic dynamics.
    Here, time-forward integration refers to simulation of changes in system state over time consecutively in discrete
time-steps, rather than solving equations that describe the whole time
evolution at once as in inter-temporal optimization.
    
    \item \textbf{Nonlinearity and tipping dynamics} WEMs should be able to capture the nonlinear dynamics that is a prerequisite for modeling climatic~\citep{lenton2008tipping} 
    and social tipping dynamics~\citep{kopp2016tipping,milkoreit2018defining} and their interactions~\citep{kriegler2009} that are not or only partially captured in ESMs and IAMs. This feature is important because the impacts of these critical dynamics are decisive for future trajectories of the Earth system in the Anthropocene, e.g. separating stabilized Earth states that allow for sustainable development from hothouse Earth states of self-amplifying global warming~\citep{Steffen2017trajectories}. 
    
    \item \textbf{Cross-scale interactions}
    Modeling approaches for investigating social-ecological or coupled human and natural system dynamics have already been developed.
    However, they usually focus on local or small-scale human-nature interactions~\citep{Schlueter2012}. 
    Therefore, such approaches need to be connected across scales and up to the planetary scale and incorporate insights from macro-level and global modeling exercises~\citep{cash2006scale}. 
    
    \item \textbf{Systematic exploration of state and parameter spaces} WEMs should allow for a comprehensive evaluation of state and parameter spaces 
    to explore the universe of accessible system trajectories 
    and to enable rigorous analyses of uncertainties and model robustness.
    Hence, they emphasize neither storylines nor optimizations 
    but focus on the exploration of the space of dynamic possibilities to gain systemic understanding.
    This principle allows for crucial Anthropocene Earth system dynamics to be investigated 
    with state-of-the-art methods from complex systems theory, 
    e.g., for measuring different aspects of stability and resilience of whole Earth system states~\citep{menck2013basin,van2016constrained,donges2017math} 
    and for understanding and quantifying planetary boundaries, safe operating spaces and their manageability and reachability as emergent system properties across scales~\citep{heitzig2016topology,kittel2017operationalization}. 

\end{enumerate}

%
%
\subsection{World-Earth models compared to existing modeling approaches of global change}
\label{sec:wem_comparison}

It is instructive to compare WEMs more explicitly than above to the two existing classes of global change models, Earth System Models and Integrated Assessment Models,
in terms of to what degree they represent biophysical, socio-metabolic/economic and socio-cultural subsystems and processes 
in the World-Earth system (Fig.\,\ref{fig:matrix}). Before discussing how model classes map to these process types, we describe the latter in more detail.

\subsubsection{Basic process taxa in World-Earth models}
\label{sec:processtaxa}

Based on the companion article by \citet{Donges2017c} that is also part of the Special Issue in Earth System Dynamics on ``Social dynamics and planetary boundaries in Earth system modeling", 
we classify processes occurring in the World-Earth system into three major taxa that represent the natural and societal spheres of the Earth system as well as their overlap (Fig.~\ref{fig:three-level}).
We give only a rough definition and abstain from defining a finer, hierarchical taxonomy,
being aware that gaining consensus among different disciplines on such a taxonomy would be unlikely,
and thus leaving the assignment of individual processes and attributes to either taxon to the respective model component developers:

\textbf{Environment (ENV; environmental, biophysical and natural processes)}
The `environment' process taxon is meant to contain biophysical or ``natural'' processes from
material subsystems of the Earth system that are not or only insignificantly shaped or designed by human societies
(e.g., atmosphere-ocean diffusion, growth of unmanaged vegetation, and maybe the decay of former waste dumps).

\textbf{Metabolism (MET; socio-metabolic and economic processes)} 
The `metabolism' process taxon is meant to contain socio-metabolic and economic processes 
from material subsystems that are designed or significantly shaped by human societies
(e.g., harvesting, afforestation, greenhouse gas emissions, waste dumping, land-use change, infrastructure building).
Social metabolism refers to the material flows in human societies and the way societies organize their exchanges of energy and materials with nature \citep{fischerkowalski1997,MartinezAlier2009}.

\textbf{Culture (CUL; socio-cultural processes)}
The `culture' process taxon is meant to contain socio-cultural processes 
from {\em immaterial} subsystems (e.g., opinion adoption, social learning, voting, policy-making) that are described in models in a way abstracted from their material basis.
Culture in its broadest definition refers to everything what people do, think and possess as members of society 
\citep[p.\,129]{Bierstedt1963}. Socio-cultural processes such as value and norm changes have been suggested to be key for understanding the deeper human dimensions of Earth systen dynamics in the Anthropocene~\citep{nyborg2016social,gerten2018deeper}

\subsubsection{Mapping model classes to Earth system processes}

Earth System Models focus on the process-detailed description of biogeophysical dynamics 
(e.g., atmosphere-ocean fluid dynamics or biogeochemistry), 
while socio-metabolic processes (e.g., economic growth, greenhouse gas emissions and land use) 
are incorporated via external forcing and socio-cultural processes 
(e.g., public opinion formation, political and institutional dynamics) 
are only considered through different scenarios 
regarding the development of exogenous socio-metabolic drivers.
Integrated Assessment Models contain a stylized description of biophysical dynamics, 
are process-detailed in the socio-metabolic/economic domains 
and are driven by narratives in the socio-cultural domain. 
In turn, WEMs should ultimately include all three domains equally. 
However, the focus of current and near-future developments in World-Earth modeling 
would likely lie on the development of a detailed description of socio-cultural processes 
because they are the ones where the least work has been done so far 
in formal Earth system modeling.

%
%
\section{The copan:CORE open World-Earth modeling framework}
\label{sec:model_description}

Here we give a short overview of the World-Earth open modeling framework copan:CORE 
that was designed following the principles given above (Sect.\,\ref{sec:blueprinting})
and is more formally described and justified in detail in the Supplementary Information. 
It enables a flexible model design around standard components and model setups 
that allows investigation of a broad set of case studies and research questions using both simple and complex models. 
Its flexibility and role-based modularization support flexible scripting by end users, interoperability and dynamic coupling with existing models, and a collaborative and structured development in larger teams.
copan:CORE is an open, code-based (rather than graphical) simulation modeling framework with a clear focus on Earth system models with endogenous human societies.
In other words, it is a tool that provides a standard way to build and run simulation models without giving preference to any particular modeling approach or theory describing human behavior and decision making and other aspects of social dynamics ~\citep{Muller-Hansen2017, schluter2017framework}.
Different model components can implement different, sometimes disputed, assumptions about human behavior and social dynamics from theories developed within different fields or schools of thought.
This allows for comparison studies in which one component is replaced by a different component modeling the same part of reality in a different way and exploring how the diverging assumptions influence the model outcomes.

All components can be developed and maintained by different model developers and flexibly composed into tailor-made models used for particular studies by again different researchers. 
The framework facilitates the integration of different types of modeling approaches.
It permits for example to combine micro-economic models (e.g., of a labor market at the level of individuals) with systems of ordinary differential equations (modeling for example a carbon cycle). Similarly, systems of implicit and explicit equations (e.g., representing a multi-sector economy) can be combined with Markov jump processes (for example representing natural hazards).
It also provides coupling capabilities to preexisting biophysical Earth system and economic integrated assessment models and thus helps to benefit from the detailed process representations embedded in these models.
Many of our design choices are based on experiences very similar to those reported in \citet{Robinson2018}, in particular regarding the iterative process of scientific modeling and the need for open code, a common language for a broader community, and a high level of consistency without losing flexibility. 
These features distinguish the copan:CORE modeling framework from existing modeling frameworks and platforms. 

A model composed with copan:CORE describes a certain part of the World-Earth system 
as consisting of a potentially varying set of {\em entities}
(``things that are'', e.g., a spot on the Earth's surface, the European Union, yourself),
which are involved in {\em processes}
(``things that happen'', e.g., vegetation growth, economic production, opinion formation)
that affect entities' {\em attributes}
(``how things are'', e.g., the spot's harvestable biomass, the EU's gross product, 
your opinion on fossil fuels, the atmosphere-ocean diffusion coefficient)
which represent the {\em variables} (including parameters) of a model.
An attribute can have a simple or complex data type, e.g.\ representing a binary variable, a whole social network, or, to facilitate interoperability and validation, a dimensional quantity with a proper physical unit.

\begin{figure}[t]
\includegraphics[width=.95 \textwidth]{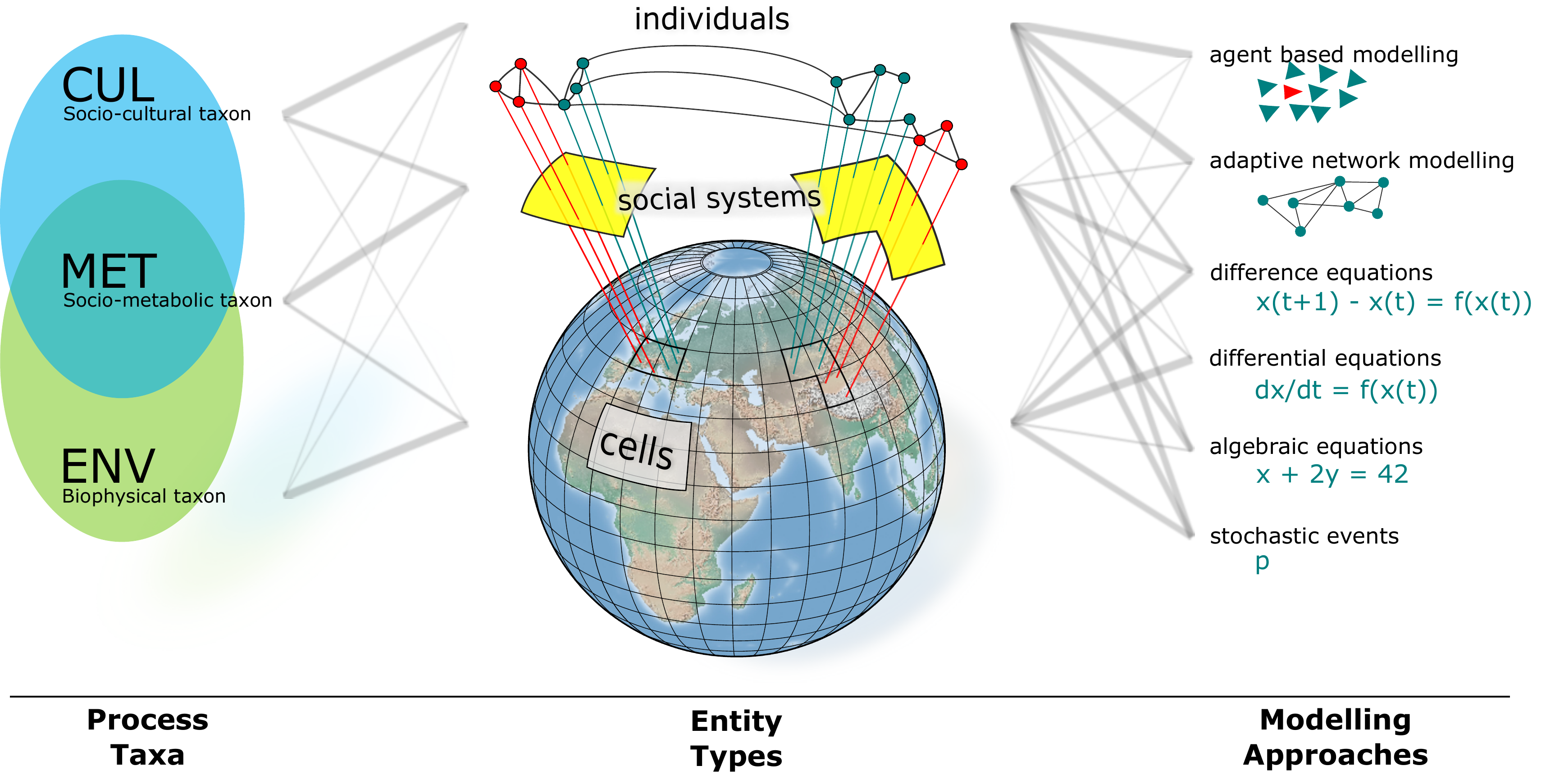}
\caption{%
    \textbf{Overview of copan:CORE open World-Earth modeling framework.} 
    The entities in copan:CORE models are classified by \textit{entity types} (e.g., grid \textit{Cell}, \textit{Social system}, \textit{Individual}, see middle column). 
    Each process belongs to either a certain entity type or a certain \textit{process taxon} (left column).
    Processes are further distinguished by formal process types (see text for a list) 
    which allow for various different \textit{modeling approaches} (right column). 
    Entity types, process taxa and process types can be freely combined with each other (grey lines).
    Thick grey lines indicate which combinations are most common.
    The copan:CORE framework allows to consistently build World-Earth models across the spectrum from stylized and globally aggregated to more complex and highly resolved in terms of space and social structure. Hence, entity types, process taxa and types may or not be present in specific models. For example, a stylized and globally aggregated model would describe the dynamics of entity types \textit{World} and \textit{Social system} and neither contain \textit{Cells} nor \textit{Individual} agents as entities.
}
\label{fig:three-level}
\end{figure}

Entities are {\em classified} by {\em entity type} (cell, social system, individual, \dots),
processes by their formal {\em process type} (see below),
and both are represented by objects in an object-oriented software design, currently using the Python programming language.
Each process and each attribute {\em belongs} to an entity type or a {\em process taxon} (environmental, socio-metabolic, socio-cultural).
Currently, the following formal process types are supported, enabling typical modeling approaches:
\begin{itemize}
\item {\em Ordinary differential equations} representing continuous time dynamics,
\item {\em Explicit} or {\em implicit algebraic equations} representing (quasi-)instantaneous reactions or equilibria, 
\item {\em Steps} in discrete time representing processes aggregated at the level of some regular time interval or for coupling with external, time-step-based models or model components, and
\item {\em Events} happening at irregular or random time points, representing e.g.\ agent-based and adaptive network components or externally generated extreme events).
\end{itemize}
Processes can be implemented either using an imperative programming style via class methods, or using symbolic expressions representing mathematical formulae.
copan:CORE's {\em modularization} and {\em role concept} distinguishes
\begin{itemize}
\item {\em Model components} developed by {\em model component developers}, implemented as subpackages of the copan:CORE software package providing interface and implementation mixin classes for entity types and process taxa,
\item {\em Models} made from these by {\em model composers}, implemented by forming final entity types and process taxa from these mixin classes,
\item {\em Studies} by {\em model end users} in the form of scripts that import, initialize and run such a model,
\item A {\em master data model} providing metadata for common variables to facilitate interoperability of model components and a common language for modelers, managed by a {\em modeling board}.
\end{itemize}
Entity types and their basic relations shipped with copan:CORE are:
\begin{itemize}
\item {\em `World',} representing the whole Earth (or some other planet). 
\item {\em `Cell',} representing a regularly or irregularly shaped spatial region used for discretising the spatial aspect 
of processes and attributes which are actually continuously distributed in space.
\item {\em `Social system'}, representing what is sometimes simply called a `society', i.e. ``an economic, social, industrial or cultural infrastructure''~\citep{wiki-society} such as a megacity, country, or the EU.
It can be interpreted as a human-designed and human-reproduced structure 
including the flows of energy, material, financial and other resources that are used to satisfy human needs and desires, 
influenced by the accessibility and usage of technology and infrastructure~\citep{fischerkowalski1997,otto2018metabolism},
and may include social institutions 
such as informal systems of norms, values and beliefs, 
and formally codified written laws and regulations, governance and organizational structures \citep{Williamson1998}.
\item {\em `Individual'}, representing a person, typically used in an network-, game-theoretic, or agent-based component. 
In contrast to certain economic modeling approaches that use ``representative'' consumers,
an `individual' in copan:CORE is not meant to represent a whole class of similar individuals 
(e.g., all the actual individuals of a certain profession)
but just one specific individual.
Still, the set of all `individuals' contained in a model 
will typically be interpreted as being a representative {\em sample} of all relevant real-world people. Each individual resides in a cell that belongs to a social system.
\end{itemize}

Fig.\,\ref{fig:three-level} illustrates these concepts.
Although there is no one-to-one correspondence between process taxa, entity types, and modeling approaches, some combinations are expected to occur more often than others, as indicated by the thicker gray connections in Fig.\,\ref{fig:three-level}.
We expect environmental (ENV) processes to deal mostly with `cells' (for local processes such as terrestrial vegetation dynamics described with spatial resolution) and `world(s)' (for global processes described without spatial resolution, e.g., the greenhouse effect)
and sometimes `social systems' (for mesoscopic processes described at the level of a social system's territory, 
e.g., the environmental diffusion and decomposition of industrial wastes).
Socio-metabolic (MET) processes will primarily deal with 
`social systems' (e.g., for processes described at national or urban level),
`cells' (for local socio-metabolic processes described with additional spatial resolution for easier coupling to natural processes)
and `world(s)' (for global socio-metabolic processes such as international trade),
and only rarely with `individuals' 
(e.g., for micro-economic model components such as consumption, investment or the job market).
Socio-cultural (CUL) processes will mostly deal with 
`individuals' (for ``micro''-level descriptions) and
`social systems' (for ``macro''-level descriptions),
and rarely `world(s)' (for international processes such as diplomacy or treaties). Other entity types such as, e.g., firms, social groups or institutions can be added to the framework if needed.

\begin{figure}[t]
\includegraphics[width=.9 \textwidth]{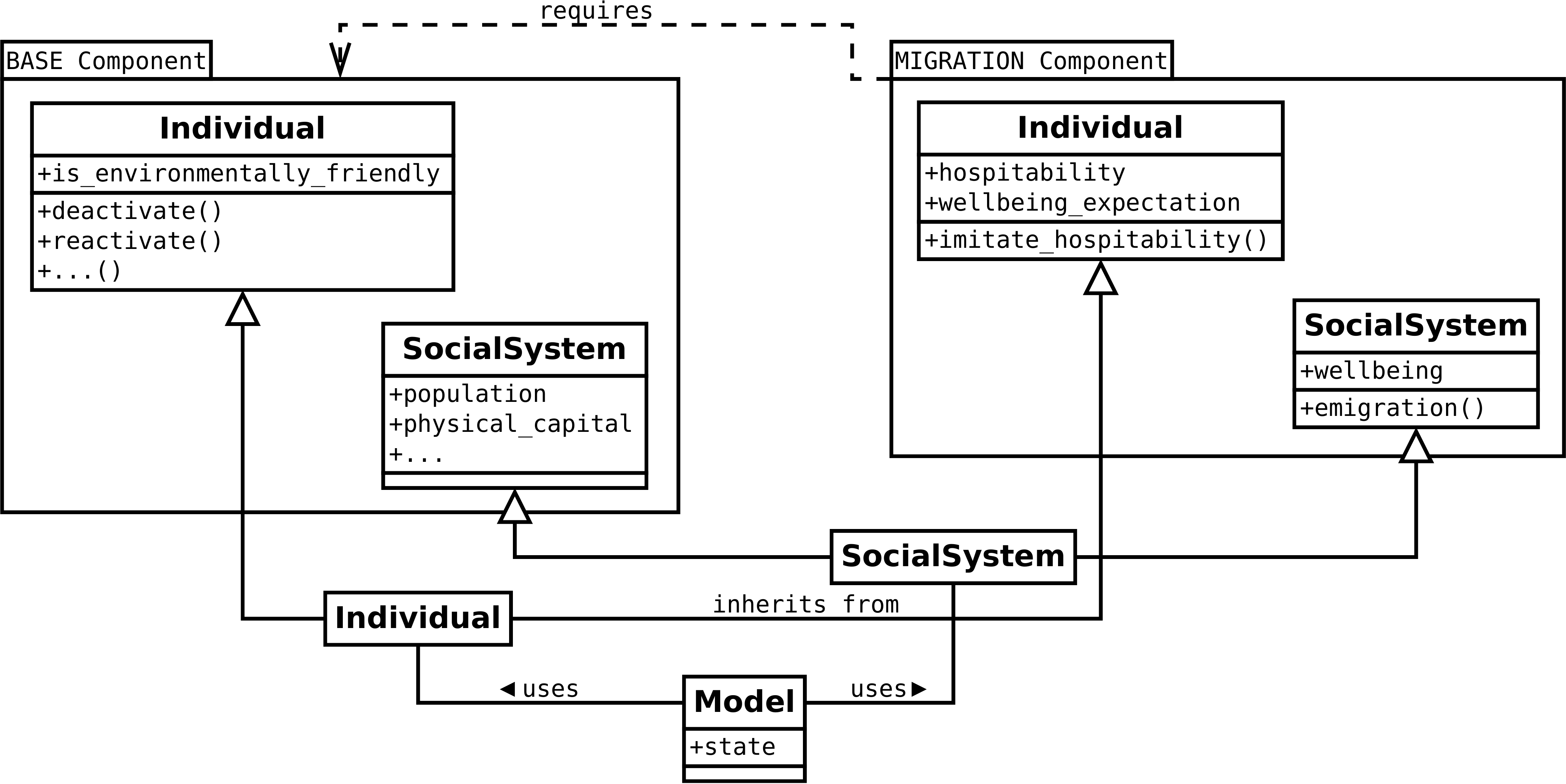}
\caption{%
    \textbf{Model composition through multiple inheritance of attributes and processes by process taxa and entity types.}
    This stylized class diagram shows how a model in copan:CORE can be composed from several model components (only two shown here, the mandatory component `base' and the fictitious component `migration') that contribute component-specific processes and attributes to the model's process taxa and entity types (only two shown here, `Individual' and `SocialSystem'). To achieve this, the classes implementing these entity types on the model level are composed via multiple inheritance (solid arrows) from their component-level counterparts (so-called `mixin' classes).  
    }
\label{fig:inheritance}
\end{figure}

\section{Influence of social dynamics in a minimum-complexity World-Earth model implemented using copan:CORE}
\label{sec:application}

In this section, we present an illustrative example of a model realized with our framework.
The example model was designed to showcase the concepts and capabilities of copan:CORE in a rather simple WEM, 
and its components were chosen so that all entity types and process taxa and most features of copan:CORE are covered.
Although most model components are somewhat plausible versions of model components that can be found in the various literatures,
the example model is intended to be a toy representation of the real world 
rather than one that could be used directly for studying concrete research questions. 
Likewise, although we show example trajectories that are based on 
parameters and initial conditions that roughly reproduce current values of real-world global aggregates 
in order to make the example as accessible as possible,
the shown time evolutions may not be interpreted as any kind of meaningful quantitative prediction or projection.

In spite of this modest goal here, 
it will become obvious from the presented scenarios
that including socio-cultural dynamics such as migration, environmental awareness, social learning, and policy making 
into more serious models of the global co-evolution of human societies and the environment
will likely make a considerable qualitative difference for their results 
and thus have significant policy implications.

\begin{figure}[t]
\centering
\includegraphics[width=.7 \textwidth, clip]{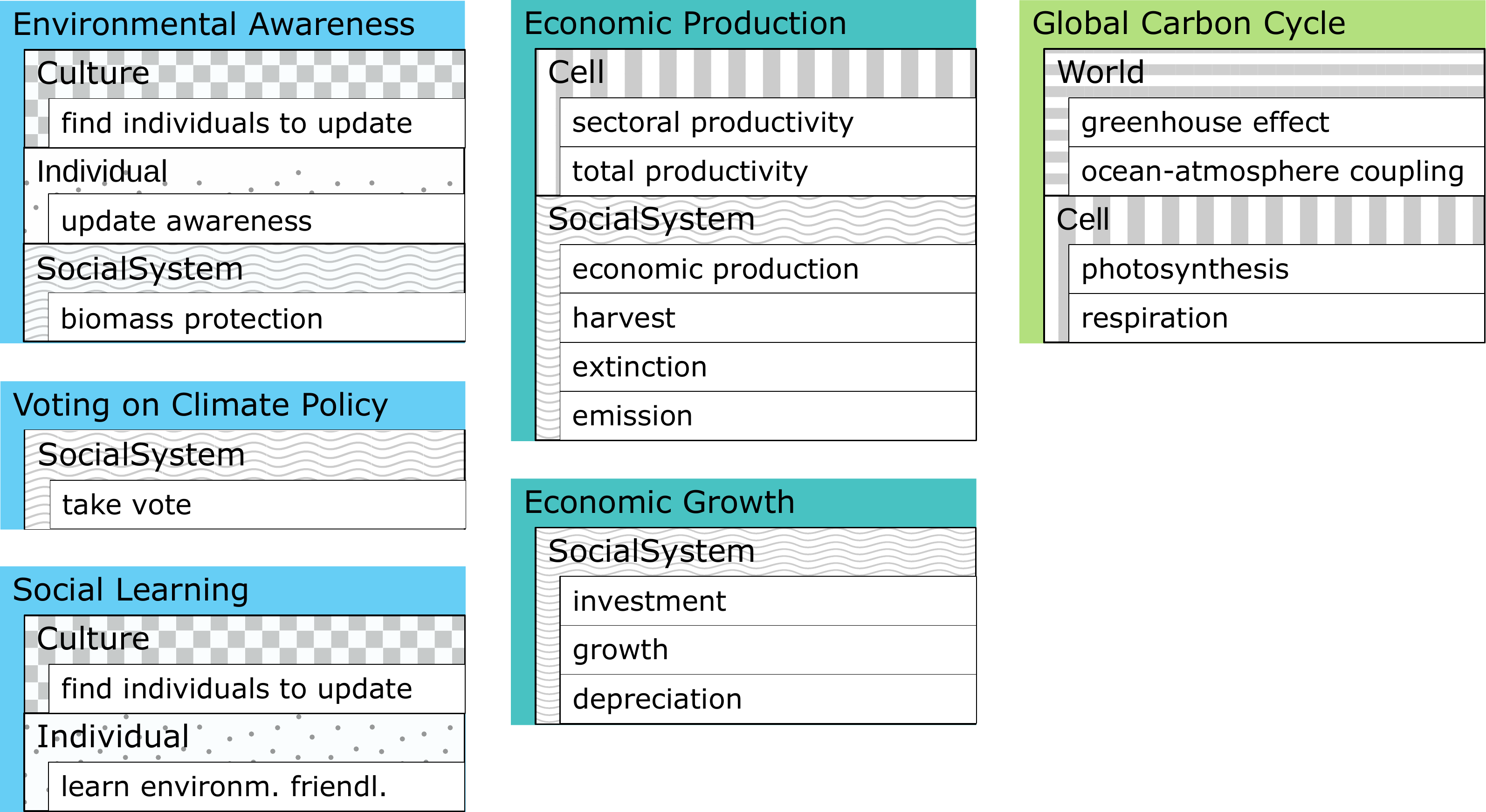}
\caption{\label{fig:examplecomponents}
    \textbf{Components, entity types, and processes of the example model.} 
    Each box represents a model component that contributes several processes
    (white bars) to different entity types and process taxa 
    (differently hashed rectangles).}
\end{figure}

The example model includes the following groups of processes:
(1) a version of the simple carbon cycle used in \citet{Nitzbon2017} \citep[based on][]{Anderies2013} coarsely spatially resolved into four heterogeneous boxes;
(2) a version of the simple economy used in \citet{Nitzbon2017} resolved into two world regions.
The fossil and biomass energy sectors are complemented by
a renewable energy sector with technological progress based on learning by doing \citep{Nagy2013}
and with human capital depreciation;
and (3) domestic voting on subsidizing renewables and banning fossil fuels
that is driven by individual environmental friendliness.
The latter results from becoming aware of environmental problems by observing the local biomass density
and diffuses through a social acquaintance network via a standard model of social learning \citep[see e.g.,][]{Holley1975}.
These processes cover all possible process taxon interactions as shown in Table~\ref{tbl:examplematrix}
and are distributed over six model components in the code as shown in Fig.\,\ref{fig:examplecomponents}.

\begin{table}
\small
\begin{tabular}{|l|l|l|l|} \hline
    $\to$   & \bf CUL   & \bf MET   & \bf ENV \\\hline
    \bf CUL & social learning, voting
                        & energy policy
                                    & environmental protection \\\hline
    \bf MET & wellbeing
                        & production, capital growth
                                    & extraction, harvest, emissions \\\hline
    \bf ENV & wellbeing, awareness
                        & resource availability
                                    & carbon cycle \\\hline
\end{tabular}

~

\caption{\label{tbl:examplematrix}
Possible classification of exemplary model processes by owning process taxon (row) 
and affected process taxon (column) \citep[following the taxonomy developed in the companion paper][]{Donges2017c}: environmental (ENV), social-metabolic (MET) and socio-cultural (CUL)
}
\end{table}

\def\pluseq{\mathrel{+\!\!=}}
We now describe the model components in detail.
As many processes add terms to variables' time derivatives,
we use the notation $\dot X \pluseq Y$ to indicate this. The effective time evolution of $X$ is then determined by the sum of the individual processes given below.

\subsection{Entity types}

The example model contains one `world' representing the planet,
two `social systems' representing the global North and South,
four `cells' representing major climate zones: `Boreal' and `Temperate' belonging to the territory of North, and `Subtropical' and `Tropical' belonging to South,
and 100 representative `individuals' per cell
which form the nodes of a fixed acquaintance network.

\subsection{Global carbon cycle}

Our carbon cycle follows a simplified version of \cite{Anderies2013} 
presented in \cite{Nitzbon2017} with a coarsely spatially resolved vegetation dynamics.
On the world level, 
an immediate greenhouse effect translates the atmospheric carbon stock $A$ 
(initially $830\,$GtC)
linearly into a mean surface air temperature 
$T = T_{\rm ref} + a (A - A_{\rm ref})$ (a process of type `explicit equation') 
with a sensitivity parameter $a = 1.5\,$K$/1000\,$GtC
and reference values $T_{\rm ref} = 287\,$K and $A_{\rm ref} = 589\,$GtC.
There is ocean-atmosphere diffusion between $A$ and the upper ocean carbon stock $M$ (initially $1065\,$GtC),
\begin{align}
    \dot A &\pluseq d (M - m A), & \dot M &\pluseq d (m A - M)
\end{align}
(processes of type `ODE'),
with a diffusion rate $d = 0.016/$yr and a solubility parameter $m = 1.5$. 
On the level of a cell $c$,
$A$ and the cell's terrestrial carbon stock $L_c$
(initially $620\,$GtC for all four $c$) 
are changed by a respiration flow $RF_c$ and a photosynthesis flow $PF_c$,
\begin{align}
    \dot A &\pluseq RF_c - PF_c, &  \dot L_c &\pluseq PF_c - RF_c.
\end{align}
The respiration rate depends linearly on temperature, 
which is expressed as a dependency on atmospheric carbon density
$A/\Sigma$, where $\Sigma = 1.5e8\,$km${}^2$ is the total land surface area, 
so that 
\begin{align}
    RF_c &= (a_0 + a_A A/\Sigma) L_c
\end{align}
with a basic rate $a_0 = 0.0298/$yr and carbon sensitivity $a_A = 3200\,$km${}^2$/GtC/yr.
The photosynthesis rate also depends linearly on temperature (and hence on $A$)
with an additional carbon fertilization factor growing concavely with $A/\Sigma$
and a space competition factor similar to a logistic equation, giving
\begin{align} 
    PF &= (l_0 + l_A A/\Sigma)\sqrt{A/\Sigma}(1 - L_c/k \Sigma_c)L_c,
\end{align}
with land area $\Sigma_c = \Sigma/4$, 
parameters
$l_0 = 34\,$km/GtC${}^{1/2}$/yr and
$l_A = 1.1e6\,$km${}^3$/GtC${}^{3/2}$/yr,
and per-area terrestrial carbon capacity $k = 25e3\,$GtC$/1.5e8\,$km$^2$.
Note that especially the linear temperature dependency and the missing water dependency
make this model rather stylized, see also \cite{LadeDonges2017}.

\subsection{Economic production}

As in \cite{Nitzbon2017}, 
economic activity consists of producing a final good $Y$ from
labour (assumed to be proportional to population $P$),
physical capital $K$
(initially $K_{North} = 4e13\,$\$, $K_{South} = 2e13\,$\$), 
and energy input flow $E$.
The latter is the sum of the outputs of three energy sectors,
fossil energy flow $E_F$, 
biomass energy flow $E_B$, 
and (other) renewable energy flow $R$. 
The process is described by a nested Leontieff/Cobb--Douglas production function for $Y$ and Cobb--Douglas production functions for $E_F,E_B,R$,
all of them here on the level of a cell $c$:
\begin{align}
    Y_c &= y_E \min(E_c, b_Y K_{Y,c}^{\kappa_Y} P_{Y,c}^{\pi_Y}), 
    & E_c &= E_{F,c} + E_{B,c} + R_c, \\
    E_{F,c} &= b_F K_{F,c}^{\kappa_F} P_{F,c}^{\pi_F} G_c^\gamma, \\
    E_{B,c} &= b_B K_{B,c}^{\kappa_B} P_{B,c}^{\pi_B} (L_c - L_c^p)^\lambda, \label{eq:EB} \\
    R_c &= b_{R,c} K_{R,c}^{\kappa_R} P_{R,c}^{\pi_R} S_s^\sigma.
\end{align}
In this, 
$y_E = 147\,$\$/GJ is the energy efficiency,
$G_c$ is the cell's fossil reserves (initially 0.4, 0.3, 0.2 and 0.1$\times 1125\,$GtC in the Boreal, Temperate, Subtropical and Tropical cells),
$L_c^p$ is the environmentally protected amount of terrestrial carbon (see below),
$S_s$ gives the renewable energy production knowledge stock of the corresponding social system $s$ (initially $2e11\,$GJ),
and $\kappa_\bullet = \pi_\bullet = \gamma = \lambda = \sigma = 2/5$ are elasticities leading to slightly increasing returns to scale.
The productivity parameters $b_\bullet$ have units that depend on the elasticities and are chosen so that initial global energy flows roughly match the observed values:
$b_F = 1.4e9\,$(GJ/yr)${}^5$/(GtC\,\$)${}^2$,
$b_B = 6.8e8\,$(GJ/yr)${}^5$/(GtC\,\$)${}^2$,
and
$b_{R,c} = $ 0.7, 0.9, 1.1 and 1.3 times the mean value $b_R = 1.75\times 10^{-11}\,$(GJ/yr)${}^5$/(GJ\,\$)${}^2$
in Boreal, Temperate, Subtropical and Tropical to reflect regional differences in solar insolation.
As in \cite{Nitzbon2017}, we assume $b_Y\gg b_B,b_F,b_R$ so that  its actual value has no influence because then $K_{Y,c}\ll K_s$ and $P_{Y,c}\ll Y_s$. 
Furthermore, $K_{\bullet,c},P_{\bullet,c}$ 
are the shares of a social system $s$'s capital $K_s$ and labour $L_s$ 
that are endogenously allocated to the production processes in cell $c$ 
so that
\begin{align}
    K_s &= \sum_{c\in s} (K_{Y,c} + K_{F,c} + K_{B,c} + K_{R,c})
\end{align}
and similarly for its population $P_s$.
The latter shares are determined on the social system level
in a general equilibrium fashion by 
equating both wages (= marginal productivity of labour) and rents (= marginal productivity of capital) in all cells and sectors,
assuming costless and immediate labour and capital mobility between all cells and sectors within each social system:
\begin{align}
    \partial y_E E_{F,c}/\partial P_{F,c} 
    \equiv \partial y_E E_{B,c}/\partial P_{B,c} 
    \equiv \partial y_E R_c/\partial P_{R,c} &\equiv w_s
\end{align} 
for all $c\in s$,
and similarly for $K_{\bullet,c}$.
The production functions and elasticities are chosen so 
that the corresponding equations can be solved analytically
(see \cite{Nitzbon2017} for details),
allowing us to first calculate a set of ``effective sector/cell productivities'' by a process of type `explicit equation' on the Cell level,
which are used to determine the labour and capital allocation weights $P_{\bullet,c}/P_s$ and $K_{\bullet,c}/K_s$,
and then calculate output $Y_s$, carbon emissions, 
and all cells' fossil and biomass extraction flows
in another process of type `explicit equation' on the social system level.
Given the latter, a second process of type `ODE' on the social system level changes the stocks $A$, $G_c$ and $L_c$ for all cells accordingly.

\subsection{Economic growth}

Again as in \cite{Nitzbon2017}, but here on the social system level,
a fixed share $i$ (here $0.244$) of economic production $Y_s$ is invested into physical capital $K_s$,
\begin{align}
    \dot K_s &\pluseq i Y_s.
\end{align}
Capital also depreciates at a rate 
that depends linearly on surface air temperature to represent damages from climate change,
\begin{align}
    \dot K_s &\pluseq - (k_0 + k_T (T - T_K)) K_s,
\end{align}
with $k_0 = 0.1/$yr,
$k_T = 0.05$/yr/K,
and $T_K = 287\,$K.
In addition,
renewable energy production knowledge $S_s$ grows proportional to its utilization via learning-by-doing,
\begin{align}
    \dot S_s &\pluseq R_s.
\end{align}
Finally, we interpret $S_s$ as a form of human capital 
that also depreciates at a constant rate 
(due to forgetting or becoming useless because of changing technology, etc.),
\begin{align}
    \dot S_s &\pluseq -\beta S_s
\end{align}
with $\beta = 0.02/$yr.
Note that unlike in \cite{Nitzbon2017}, we consider populations to be constant at 
$P_{North}s = 1.5e9$ and 
$P_{South}s = 4.5e9$ 
to avoid the complexities of a wellbeing-driven population dynamics component (which could however be implemented in the same way as in \cite{Nitzbon2017} on the social system level).

\subsection{Environmental awareness}

On the level of the `Culture' process taxon, an ``awareness updating'' process of type `event' occurs at random time points
with a constant rate (i.e., as a Poisson process, here with rate 4/yr),
representing times at which many people become aware of the state of the environment,
e.g., because of notable environmental events. 
At each such time point, each individual independently updates
her environmental friendliness (a Boolean variable)
with a certain probability.
When $i$ updates,
she switches from ``false'' to ``true'' with a probability $\psi^+$ depending on the terrestrial carbon density in her cell $c$,
$TCD_c = L_c/\Sigma_c$, given by
\begin{align}
    \psi^+ &= \exp(- TCD_c / TCD^\bot),
\end{align}
and switches from ``true'' to ``false'' with a probability
\begin{align}
    \psi^- &= 1 - \exp(- TCD_c / TCD^\top),
\end{align}
where $TCD^\bot = 1e{-}5$ 
and $TCD^\top = 4e{-}5$ 
are sensitivity parameters
with $TCD^\bot < TCD^\top$ to generate hysteresis behaviour.
As a consequence, 
a fraction $L_c^p$ of the terrestrial carbon $L_c$ is protected from harvesting for economic production.
This fraction is proportional to the cell's social system's population share 
represented by those individuals $i$ which are environmentally friendly.
The initial share of environmentally friendly individuals will be varied in the bifurcation analysis below.

\subsection{Social learning}

Similarly, on the Culture level, ``social learning'' events occur at random time points
with a constant rate (here 4/yr),
representing times at which the state of the environment becomes a main topic in the public debate. 
At each such time point, each individual $i$ independently compares 
her environment with that of a randomly chosen acquaintance $j$
with a certain fixed probability (here $1/10$).
$j$ then convinces $i$ to copy $j$'s environmental friendliness 
with a probability $\psi$ that depends via a sigmoidal function
on the difference-in-logs between both home cells' terrestrial carbon densities,
\begin{align}
    \psi &= 1/2 + \arctan (\pi \phi' (\log TCD_j - \log TCD_i - \log \rho')) / \pi,
\end{align}
where $\phi' = 1$ 
and $\rho' = 1$ 
are slope and offset parameters.
The underlying social network is a block model network in which each individual is on average linked to 10 randomly chosen others: 5 in the same cell, 3.5 in the other cell of the same social system, and 1.5 in the other social system.

\subsection{Voting on climate policy}

Each (of the two) social systems performs general elections at regular time intervals
(hence implemented as a process of type `step', here every 4 years)
which may lead to the introduction or termination of climate policies.
If at the time $t$ of the election, 
more than a certain threshold (here $1/2$) of the population is environmentally friendly,
both a subsidy for renewables (here $50\,$\$/GJ) is introduced and use of fossils is banned.
This leads to a shift in the energy price equilibrium that determines the energy sector's allocation of labour and capital,
which then reads
\begin{align*}
    &\text{marginal production cost of biomass energy} \\
    &= \text{marginal production cost of renewable energy} - \text{renewable subsidy}.
\end{align*}
Conversely, if these policies are already in place 
but the environmentally friendly population share is below some other thresholds (here as well $1/2$),
these policies are terminated. 

~

Note that we have chosen to model awareness-formation and social learning in an agent-based fashion here mainly to illustrate that such an approach can be easily combined with other approaches in copan:CORE, not because we want to claim that an agent-based approach is the most suitable here. Indeed, one may well want to replace these two agent-based model components by equation-based versions which approximate their behaviour in terms of macroscopic quantities (e.g. as in \citet{Wiedermann2015}), and because of the modular design of copan:CORE, this can easily be done and the two model versions could be compared (still, this is beyond the scope of this paper).

\subsection{Results}
\begin{figure}[t]
\includegraphics[width=.75 \textwidth]{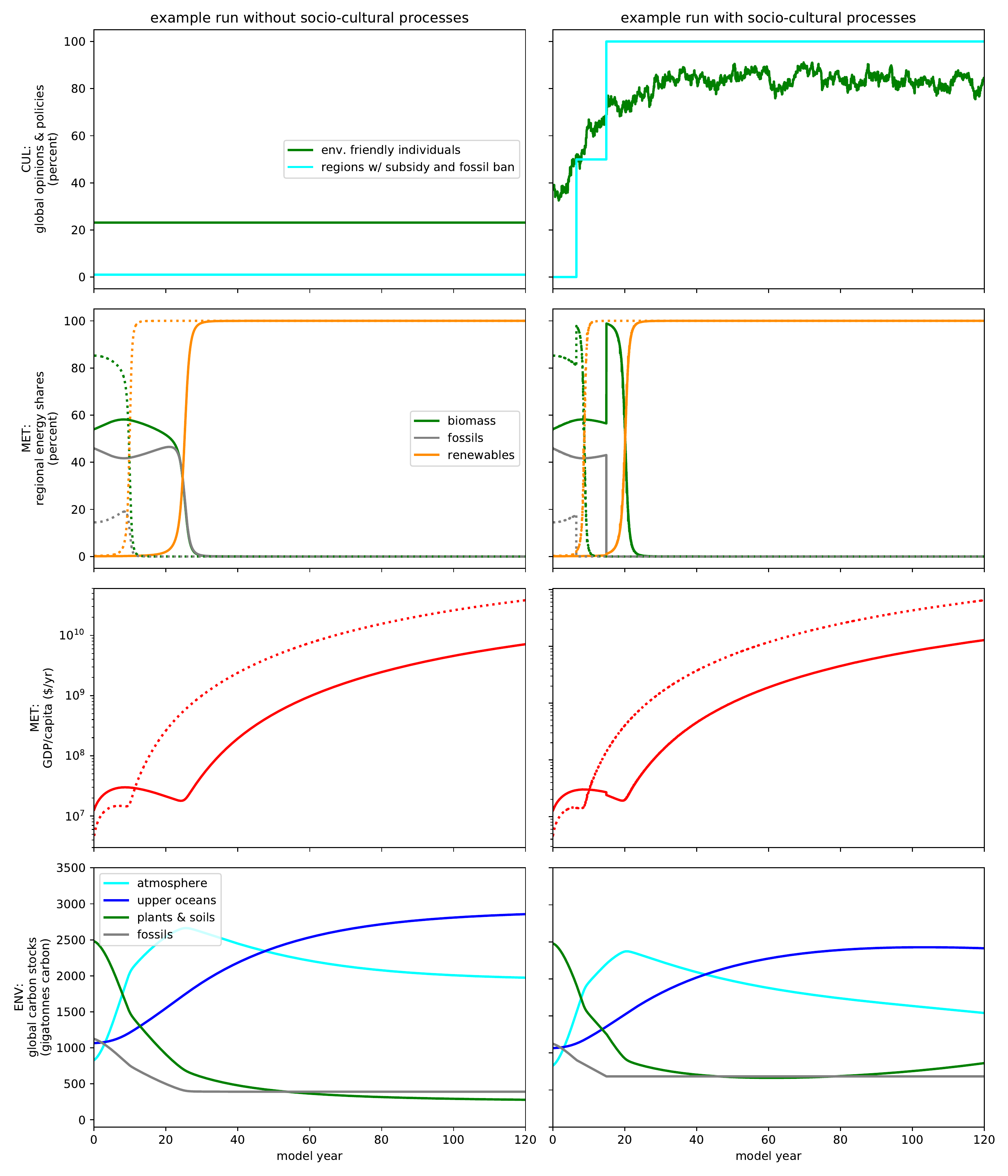}
\caption{\label{fig:results}
    \textbf{Two runs from a World-Earth model example,} one without (left) and one with (right) the socio-cultural processes of
    environmental awareness, social learning, and voting included,
    showing different transient (and asymptotic) behavior. The top row shows variables related to the cultural process taxon, the center row those related to the metabolic process taxon and the bottom row those related to the environmental process taxon. 
    Green/orange/cyan/blue/gray lines correspond to variables related to terrestrial carbon /  renewables / atmospheric carbon / ocean carbon / fossils. In the middle two panels, dashed lines belong to the `South', solid lines to the `North'.}
\end{figure}

\begin{figure}[t]
\includegraphics[width=.5 \textwidth,]{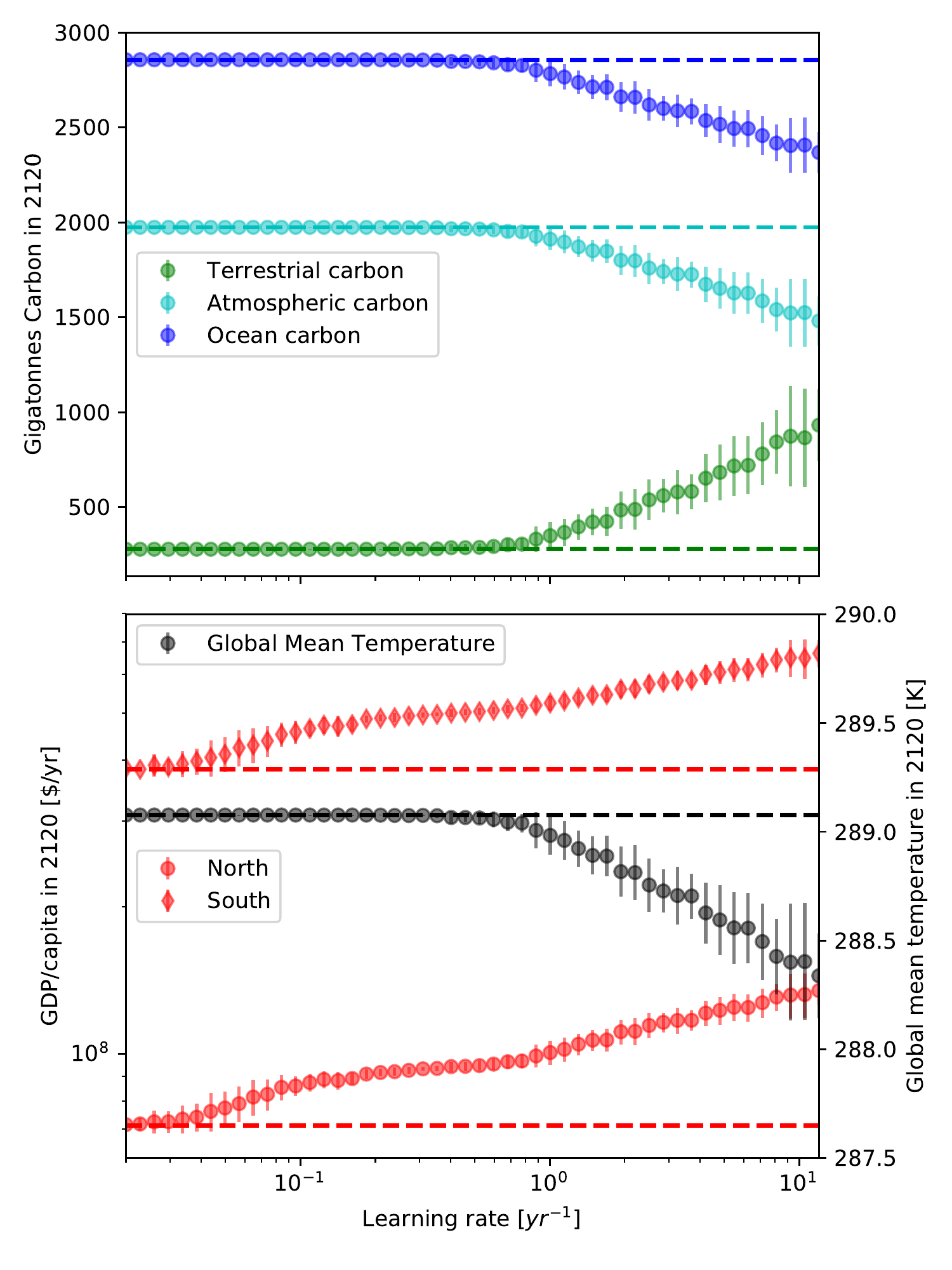}
\caption{Dependency of some selected variables after 120 model years on the learning rate of environmental awareness. Scatter points denote (the average over 50) simulations with social processes and error bars denote one standard deviation for each choice of learning rate. Dashed lines indicate the corresponding values for a simulation without social processes. The top panel shows the three environmental (non-fossil) carbon stocks, the bottom panel shows the GDP per capita in the two social systems as well as the global mean temperature.}
\label{fig:bifurcation}
\end{figure}

In order to show in particular what effect the inclusion of socio-cultural processes into WEMs can have on their results,  we compare two representative hundred-year runs of the example model described above, one without the socio-cultural processes environmental awareness, social learning, and voting (left panel of Fig.~\ref{fig:results}), and another with these processes included (right panel of Fig.~\ref{fig:results}). Both runs start in model year 0 from the same initial conditions and use the same parameters which were chosen to roughly reflect real-world global aggregates of the year 2000 (see above). For the simulation without social processes (left panel of Fig.~\ref{fig:results}) both social systems ('North' in solid and 'South' in dashed lines) initially rely on fossil energy in order to meet their energy needs, thus causing a rise in in atmospheric and ocean carbon and a decline in fossil carbon stocks. Similarly both social systems initially rely heavily on energy from biomass, with the consequence of a reduction in terrestrial carbon. Due to the technology becoming competitive, the South changes its energy production to renewable energy comparatively early in the simulation, resulting in a fast fading out of biomass and fossils as an energy source. 
Due to its larger fossil reserves and lower solar insolation, the North takes two decades longer to make this switch.
However, this delay in the North causes high atmospheric carbon, hence a high global mean temperature, which due to our oversimplified vegetation model makes the terrestrial carbon stock decline further even after biomass has been phased out as an energy source as well, recovering only much later (not shown). 
In both social systems, economic growth declines until the switch, then boosts and later declines again since neither population nor total factor productivity grow in our model. Once the South switches to renewables, they hence overtake the North, and this reversed inequality is then sustained since our model includes no trade, knowledge spillovers, migration or other direct interaction which would lead to economic convergence. Certainly, such results are not in itself realistic (as this model does not intend to be) or transferable to real-world application. Future WEM's, therefore, should include such processes beyond pure economic ones in order to properly capture real-World-Earth dynamics; see the Supplementary Information for some corresponding extensions of this model.                    

If social processes are considered, we obtain qualitatively similar, but quantitatively different trajectories, e.g. in the right panel of Fig.~\ref{fig:results}, where we assume initially 40\% are environmentally friendly. As before, both social systems initially rely on energy produced from fossils and biomass, but as biomass reduces terrestrial carbon density, environmental awareness makes some people environmentally friendly and this spreads via social learning. Once half of the population is environmentally friendly, the next elections in that social system bring a fossil ban and subsidies for renewables. This causes a slightly earlier switch to renewables than before, especially in the North (dashed lines in Fig.~\ref{fig:results}). This ultimately results in lower atmospheric and ocean carbon stocks, lower peak temperatures, less cumulative use of fossil fuels, and a much faster recovery of terrestrial carbon.

copan:CORE further allows for a systematic investigation of the influence of individual parameter on the outcome of the simulation (e.g. along the lines of a bifurcation analysis). As an illustration of such an analysis we now vary the learning rate from $1/50$yr
(less than once in a generation) to $12/$yr 
(once every month) and compute the carbon stocks as well as the GDP per capita and the global mean temperature in model year 120 for an ensemble of 50 simulations per learning rate (Fig.~\ref{fig:bifurcation}) and the same initial conditions for all runs (we thus do not test for a possible multistability of the system).

For learning rates lower than $1/$yr (slow learning) the carbon stocks as well as the global mean temperature align well for the two simulation setups, i.e, the one with (scatter points) and without social processes (dashed lines). In contrast, for learning rates larger than $1/$yr (faster learning) the individuals become more capable of assessing the consequences of their behaviour (in our case extensive biomass use) before the system has reached a state with low terrestrial and high atmospheric and ocean carbon stocks. As such, increasing the learning rate also causes an increase in the terrestrial carbon stock combined with a decrease of the atmospheric and ocean carbon stocks (in model year 120). This behaviour is also reflected in the global mean temperature which decreases as the learning rate increases. Hence, with respect to the environment social learning only has a positive effect if it happens at sufficiently high rate (around once to more than once a year). It remains to note that learning rates have in the past already been shown to have a profound impact on the state and dynamics of a coupled socio-ecological system, a feature that is recovered in our simple WEM as well~\citep{Wiedermann2015, Auer2015, barfuss2017sustainable}.

The metabolic variable GDP per capita interestingly already increases much earlier (i.e., for much lower learning rates than $1/$yr) as compared to the changes in the environmental variables. This implies that for our specific WEM social processes generally seem to foster the economy regardless of their actual rate. Furthermore we observe that the South shows an approximately 3 times higher GDP per capita than the North, which is caused by the earlier switch to renewable energies in that social system (see third row of Fig.~\ref{fig:results}). As already stated above, note again, that these results are not intended as a realistic projection of future trajectories of the Earth System, but are discussed here to showcase the capabilities of the copan:CORE framework.

Using the \textit{pycopancore} reference implementation, running the above two simulations (Figure~\ref{fig:results}) took 140 seconds (without socio-cultural processes) and 290 seconds (including socio-cultural processes) on an E5-2690 CPU at 2.60\,GHz. Since further performance improvements are desirable to support Monte-Carlo simulations, we aim at a community-supported development of an alternative, more production-oriented implementation in the C++ language.

\conclusions
\label{sec:conclusions}

In this paper, we presented a simulation modeling framework 
that aims at facilitating the implementation and analysis of  World-Earth (or planetary social-ecological) models.
It follows a modular design such that various model components can be combined in a plug-and-play fashion 
to easily explore the influence of specific processes 
or the effect of competing theories of social dynamics from different schools of thought~\citep{schluter2017framework} 
on the co-evolutionary trajectory of the system. 
The model components describe fine-grained yet meaningfully defined subsystems 
of the social and environmental domains of the Earth system 
and thus enable the combination of modeling approaches from the natural and social sciences.
In the modeling framework, different entities such as geographic cells, individual humans, and social systems are represented 
and their attributes are shaped by environmental, socio-metabolic, and socio-cultural processes.
The mathematical types of processes that can be implemented in the modeling framework 
range from ordinary differential and algebraic equations to deterministic and stochastic events.
Due to its flexibility, the model framework can be used to analyze interactions at and between various scales 
-- from local to regional and global.

The current version of the copan:CORE open modeling framework includes a number of tentative model components 
implementing, e.g., basic economic, climatic, biological, demographic and social network dynamics. 
However, to use the modeling framework for rigorous scientific analyses, 
these components have to be refined, their details have to be spelled out, 
and new components have to be developed 
that capture processes with crucial influence on World-Earth co-evolutionary dynamics.
For this purpose, various modeling approaches from the social sciences are available to be applied 
to develop comprehensive representations of such socio-metabolic and socio-cultural processes \citep[][and references therein]{Muller-Hansen2017}.
For example, hierarchical adaptive network approaches could be used to model 
the development of social groups, institutions and organizations spanning local to global scales 
or the interaction of economic sectors via resource, energy and information flows~\citep{gross2008adaptive,Donges2017a,geier2019physics}.

Making such an endeavor prosper requires the collection and synthesis of knowledge from various disciplines. 
The modular approach of the copan:CORE open modeling framework 
supports well-founded development of single model components, helps to integrate various processes and allows to analyze their interplay.
We therefore call upon the interdisciplinary social-ecological modeling community and beyond
to participate in further model and application development to facilitate ``whole'' Earth system analysis of the Anthropocene.


\codeavailability{A Python 3.6.x implementation of the copan:CORE open World-Earth modeling framework, its detailed documentation and the World-Earth model example are available at \url{https://github.com/pik-copan/pycopancore}.} 













\competinginterests{The authors declare no competing interests.} 


\begin{acknowledgements}
This work has been carried out within the framework of 
PIK's flagship project on Coevolutionary Pathways in the Earth system (copan, \url{www.pik-potsdam.de/copan}). 
We are grateful for financial support by the Stordalen Foundation via the Planetary Boundary Research Network (PB.net), 
the Earth League's EarthDoc programme, 
the Leibniz Association (project DominoES),
the European Research Council (ERC advanced grant project ERA), 
and the German Federal Ministry of Education and Research (BMBF, project CoNDyNet). 
We acknowledge additional support by the Heinrich B\"oll Foundation (WB), 
the Foundation of German Business (JJK),
the Episcopal Scholarship Foundation Cusanuswerk (JK)
and DFG/FAPESP (IRTG 1740/TRP 2015/50122-0, FMH).
The European Regional Development Fund, BMBF, and the Land Brandenburg supported this project 
by providing resources on the high-performance computer system at the Potsdam Institute for Climate Impact Research. We thank the participants of the three LOOPS workshops (\url{www.pik-potsdam.de/loops}) in Kloster Chorin (2014), Southampton (2015) and Potsdam (2017) for discussions that provided highly valuable insights for conceptualizing World-Earth modeling and the development of the copan:CORE open simulation modeling framework.
\end{acknowledgements}

\bibliographystyle{copernicus}
\bibliography{copancore_literature,literature_jona,literature_jobst,copancore_si}

\end{document}


\maketitle

\def\pluseq{\mathrel{+\!\!=}}

%
%
\section{The copan:CORE open World-Earth modeling framework -- Details}
\label{sec:si_model_description}

In this section, we present the World-Earth modeling framework copan:CORE 
that was designed following the principles given in Sect.\ 1 of the main text in more detail than in Sect.\ 2 of the main text. 
Here we describe our framework on three levels,
starting with the abstract level independent of any software (Sects.\,\ref{sec:core_abstract} and \ref{sec:entitytypes}),
then describing the software design independent of any programming language (Sect.\,\ref{sec:core_software}),
and finally presenting details of our reference implementation in the Python language (Sect.\,\ref{sec:core_python}).

In summary, copan:CORE enables a flexible model design around standard components and model setups 
that allows investigation of a broad set of case studies and research questions (Fig.\,\ref{fig:si_three-level}). 
Its flexibility and role-based modularization are realized within an object-oriented software design and support flexible scripting by end users and interoperability and dynamic coupling with existing models 
-- e.g., the terrestrial vegetation model {\em LPJmL} working on the cell level~\citep{bondeau2007modelling} or
other Earth system models or integrated assessment models based on time-forward integration (rather than intertemporal optimization) such as {\em IMAGE}~\citep{vanvuuren2015pathways}. 
On the level of model infrastructure, 
a careful documentation 
and software versioning via the `git' versioning system 
aim to support collaborative and structured development in large teams using copan:CORE.

\subsection{Features of the copan:CORE modeling framework}


The {\em copan:CORE World-Earth modeling framework} presented in this paper is a code-based (rather than graphical) simulation modeling framework with a clear focus on Earth system models with complex human societies.
It was developed within the flagship project `copan -- coevolutionary pathways' and will form the core of its further model development, which explains the naming.
Similar to the common definition of `software framework', we define a `(simulation) modeling framework' as a tool that provides a standard way to build and run simulation models.

We have designed copan:CORE
to meet the special requirements for model development in the context of Earth system analysis:
First, the framework's modular organization combines processes into model components.
Different components can implement different, sometimes disputed, assumptions about human behavior and social dynamics from theories developed within different fields or schools of thought.
This allows for comparison studies in which one component is replaced by a different component modeling the same part of reality in a different way and exploring how the diverging assumptions influence the model outcomes.
All components can be developed and maintained by different model developers and flexibly composed into tailor-made models used for particular studies by again different researchers. 
Second, our framework provides coupling capabilities to preexisting biophysical Earth system and economic integrated assessment models and thus helps to benefit from the knowledge of the detailed processes embedded in these models.

Finally, copan:CORE facilitates the integration of different types of modeling techniques.
It permits for example to combine agent-based models (e.g., of a labor market at the micro-level of individuals) with systems of ordinary differential equations (modeling for example a carbon cycle). Similarly, systems of implicit and explicit equations (e.g., representing a multi-sector economy) can be combined with Markov jump processes (for example representing economic and environmental shocks).

These features distinguish the copan:CORE modeling framework from existing modeling frameworks and platforms. Before we continue with a more detailed description of the modeling framework, we go back to the underlying design principles of WEMs that guided the development of copan:CORE.

\subsection{Abstract structure}
\label{sec:core_abstract}

This section describes the abstract structure of models that can be developed with copan:CORE and gives rationales for our design choices, many of which are based on experiences very similar to those reported in \citet{Robinson2018}, in particular regarding the iterative process of scientific modeling and the need for open code, a common language, and a high level of consistency without losing flexibility. 

\paragraph{Entities, processes, attributes}
A model composed with copan:CORE describes a certain part of the World-Earth system 
as consisting of a potentially large set (that may change over model time) 
of sufficiently well-distinguishable {\em entities}
(``things that are'', e.g., a spot on the Earth's surface, the European Union [EU], yourself).
Entities are involved in
a number of sufficiently well-distinguishable {\em processes}
(``things that happen'', e.g., vegetation growth, economic production, opinion formation).
Processes in turn affect one or more {\em attributes}
(``how things are'', e.g., the spot's harvestable biomass, the EU's gross product, 
your opinion on fossil fuels, the atmosphere-ocean diffusion coefficient).
During a model run, entities may come into existence 
(individuals may be born, social systems may merge into larger ones or fractionate), 
cease to exist (individuals may die, social systems may collapse),
or may even be ``reactivated'' (e.g., an occupied country may regain independence).

\textbf{Rationale.} While for some aspects of reality an ontological distinction between entities, attributes of entities, and processes might be ambiguous, it corresponds very well to both the distinction of nouns, adjectives, and verbs in natural languages, and to the concepts of objects, object attributes, and methods in object-oriented programming. 

\begin{figure}[t]
\includegraphics[width=.95 \textwidth]{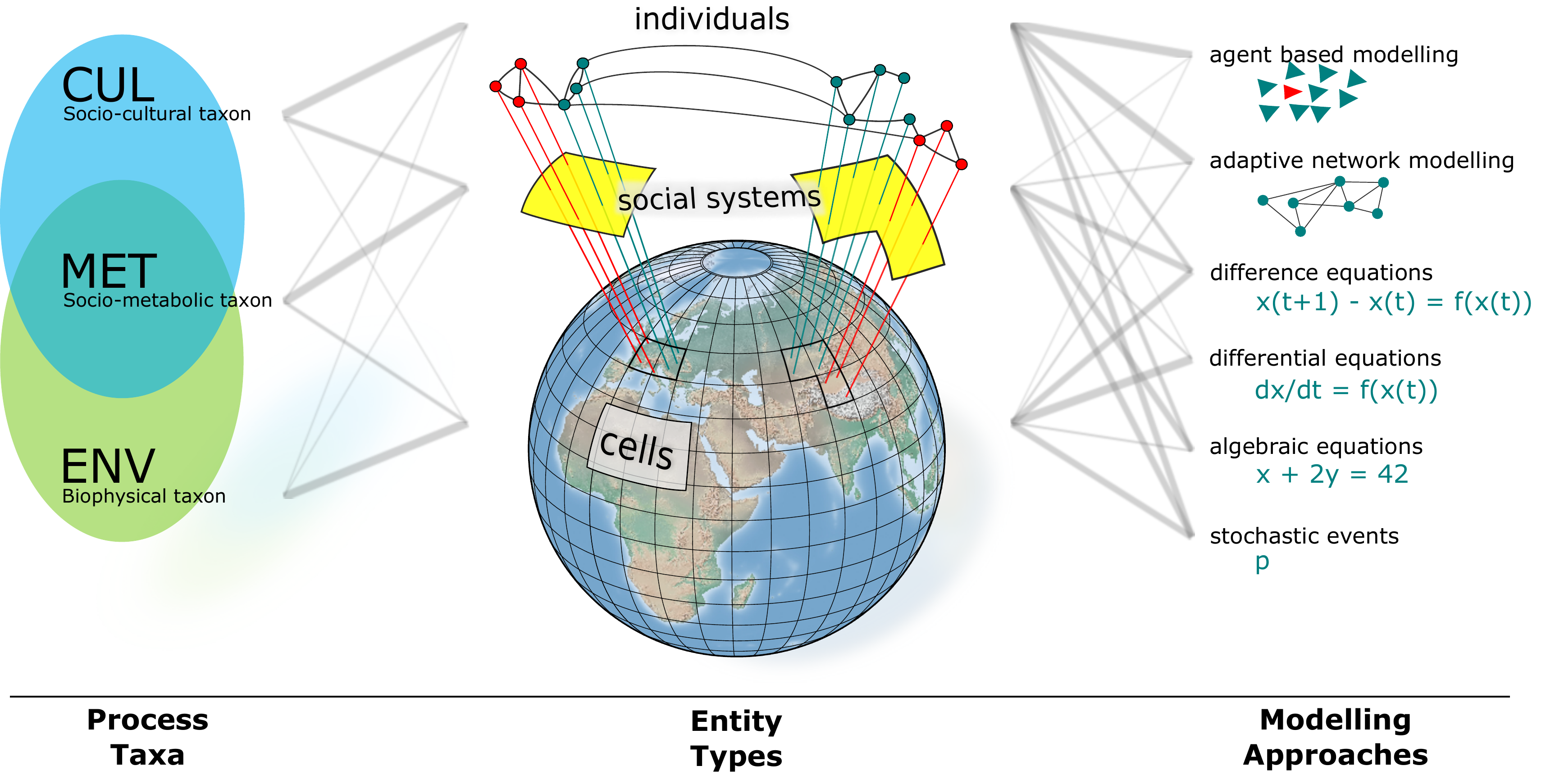}
\caption{%
    \textbf{Overview of copan:CORE open World-Earth modeling framework.} 
    The entities in copan:CORE models are classified by \textit{entity types} (e.g., grid cell, social system, individual, see middle column). 
    Each process belongs to either a certain entity type or a certain \textit{process taxon} (left column).
    Processes are further distinguished by formal process types (see text for a list) 
    which allow for various different \textit{modeling approaches} (right column). 
    Entity types, process taxa and process types can be freely combined with each other (grey lines).
    Thick grey lines indicate which combinations are most common.
}
\label{fig:si_three-level}
\end{figure}

\paragraph{Entity types, process taxa, process types}
copan:CORE classifies entities by {\em entity types}
(``kinds of things that are'', e.g., spatial grid cell, social system, individual),
and allows to group (some or all) processes into {\em process taxa}
(e.g., environmental, social-metabolic, socio-cultural).
Each process and each attribute
{\em belongs to} either a certain entity type or a certain process taxon.
We deliberately do not specify criteria for deciding where processes belong
since this is in part a question of style and academic discipline
and there will inevitably be examples where this choice appears to be quite arbitrary
and will affect only the model's description, implementation, and maybe its running time, but not its results.

Similarly, attributes may be modeled as belonging to some entity type
(e.g., `total population' might be modeled as an attribute of the `social system' entity type)
or to some process taxon
(e.g., `atmosphere-ocean diffusion coefficient' might be modeled as an attribute of the `environment' process taxon).
We suggest to model most quantities as entity type attributes
and model only those quantities as process taxon attributes which represent global constants.

Independently of where processes belong to,
they are also distinguished by their formal {\em process type,}
corresponding to different mathematical modeling and simulation/solving techniques:
\begin{itemize}
\item continuous dynamics given by ordinary differential equations,
\item (quasi-)instantaneous reactions given by algebraic equations (e.g., for describing economic equilibria), 
\item steps in discrete time (e.g., for processes aggregated at annual level 
    or for coupling with external, time-step-based models or model components), or
\item events happening at irregular or random time points 
    (e.g., for agent-based and adaptive network components or externally generated extreme events).
\end{itemize}
the latter two potentially have probabilistic effects. Later versions will also include support for stochastic differential equations or other forms of time-continuous noise,
currently noise can only be modeled via time-discretized steps.
Similarly, attributes have {\em data types}
(mostly physical or socio-economic simple quantities of various {\em dimensions} and {\em units,}
but also more complex data types such as references or networks).

Fig.\,\ref{fig:si_three-level} summarizes our basic process taxa and entity types,
their typical connections, and the corresponding typical modeling approaches 
(which in turn are related but not equal to certain formal process types, not shown in the figure).
Sects.\,\ref{sec:entitytypes} describes them in detail.

\textbf{Rationale.}
When talking about processes, people from very different backgrounds widely use a subject-verb-object sentence structure
even when the subject is not a conscious being and the described action is not deliberate
(e.g., ``the oceans take up carbon from the atmosphere'').
copan:CORE therefore allows modelers to treat some processes 
as if they were ``done by'' a certain entity (the ``subject'' of the process)
``to'' itself and/or certain other entities (the ``objects'' of the process).
Other processes for which there appears to be no natural candidate entity to serve as the ``subject''
can be treated as if they are happening ``inside'' or ``on'' some larger entity 
that contains or otherwise supports all actually involved entities.
In both cases, the process is treated as belonging to some entity type.
Still other processes such as multilateral trade may best be treated as not belonging to a single entity
and can thus be modeled as belonging to some process taxon.

A twofold classification of processes according to both ownership and formal process type is necessary since there is no one-to-one relationship between the two, as the grey lines in Fig.\,\ref{fig:si_three-level} indicate. E.g., processes from all three taxa may be represented by ODEs or via stochastic events, and all shown entity types can own regular time stepped processes.

\paragraph{Modularization, model components, user roles}
copan:CORE aims at supporting a plug-and-play approach to modeling
and a corresponding division of labour between several user groups (or {\em roles})
by dividing the overall model-based research workflow into several tasks.
As a consequence, we formally distinguish between model components and (composed) models.

A {\em model component} specifies 
(i) a meaningful collection of processes that belong so closely together
that it would not make much sense to include some of them without the others into a model
(e.g., plants' photosynthesis and respiration),
(ii) the entity attributes that those processes deal with,
referring to attributes listed in the master data model whenever possible,
(iii) which existing (or, if really necessary, additional) entity types and process taxa 
these processes and attributes belong to.
A {\em model} specifies
(i) which model components to use,
(ii) if necessary, which components are allowed to overrule parts of which other components
(iii) if necessary, any {\em attribute identities}, i.e., 
whether some generally distinct attributes should be considered to be the same thing in this model
(e.g., in a complex model, the attribute `harvestable biomass' used by an `energy sector' component as input
may need to be distinguished from the attribute `total vegetation' governed by a `vegetation dynamics' component,
but a simple model that has no `land use' component that governs their relationship may want to identify the two).

The suggested workflow is then this:
\begin{itemize}
\item If there is already a model that fits your research question, 
        use that one in your study (role: {\em model end user}).
\item If not, decide what model components the question at hand needs.
  \begin{itemize}
  \item If all components exist, compose a new model from them (role: {\em model composer}).
  \item If not, design and implement missing model components (role: {\em model component developer}).
        If some required entity attributes are not yet in the master data model (Sect.\,\ref{sec:master_data_model}), 
        add them to your component.
        Suggest well-tested entity attributes, entity types, or model components 
        to be included in the copan:CORE community's master data model or master component repository
        ({\em modeling board members} will then review them).
  \end{itemize}
\end{itemize}

\textbf{Rationale.} 
Although in smaller teams, one and the same person may act in all of the above roles, the proposed role concept helps structuring the code occurring in a model-based analysis into parts needed and maintained by different roles, a prerequisite for collaborative modeling, especially across several teams.

The additional concept of model components (in addition to entity types and taxa) is necessary since processes which belong together from a logical point of view and are hence likely to be modeled by the same person or team may still most naturally be seen as being owned by different entity types, and at the same time developers from several teams may be needed to model all the processes of some entity type.

\paragraph{Master data model and master component repository} \label{sec:master_data_model}
The {\em master data model} defines entity types, process taxa, attributes, and physical dimensions and units 
which the modeling board members deem 
(i)~likely to occur in many different models or model components
and (ii)~sufficiently well-defined and well-named
(in particular, specific enough to avoid most ambiguities but avoiding a too discipline-specific language).
Users are free to define additional attributes in their components but are encouraged to use those from the master data model or suggest new attributes for it.

The {\em master component repository} contains model components which the modeling board members
deem likely to be useful for many different models, sufficiently mature and well-tested,
and indecomposable into more suitable smaller components.
Users are free to distribute additional components not yet in the repository.

\textbf{Rationale.}
Poorly harmonized data models are a major obstacle for comparing or coupling simulation models. Still, a perfectly strict harmonization policy that would require the prior approval of every new attribute or component would inhibit fast prototyping and agile development. This is why the above two catalogs and the corresponding role were introduced. 

\paragraph{All attributes are treated as variables with metadata}
Although many models make an explicit distinction between ``endogenous'' and ``exogenous'' variables and ``parameters'',
our modular approach requires us to treat all relevant entity type or process taxon attributes a priori in the same way,
calling them {\em variables}
whether or not they turn out to be constant during a model run or are used for a bifurcation analysis in a study.

A variable's specification contains {\em metadata} such as
a common language label and description, 
possibly including references to external metadata catalogs
such as the Climate and Forecast Conventions' Standard Names \citep{CFStandardNames} for climate-related quantities or the World Bank’s CETS list of socio-economic indicators \citep{CETS},
a mathematical symbol, 
its level of measurement or scale of measure (ratio, interval, ordinal, or nominal),
its physical or socio-economic dimension and default unit (if possible following some established standard),
its default (constant or initial) value and range of possible values.

\textbf{Rationale.}
The common treatment of variables and parameters is necessary because a quantity that one model component uses as an exogenous parameter that will not be changed by this component
will often be an endogenous variable of another component,
and it is not known to a model component developer which of the quantities she deals with 
will turn out to be endogenous variables or exogenous parameters of a model or study that uses this component.
Well-specified metadata are essential for collaborative modeling to avoid hard-to-detect mistakes involving different units or deviating definitions.

\subsection{Basic entity types}
\label{sec:entitytypes}

\begin{figure}[t]
\includegraphics[width=.95\textwidth]{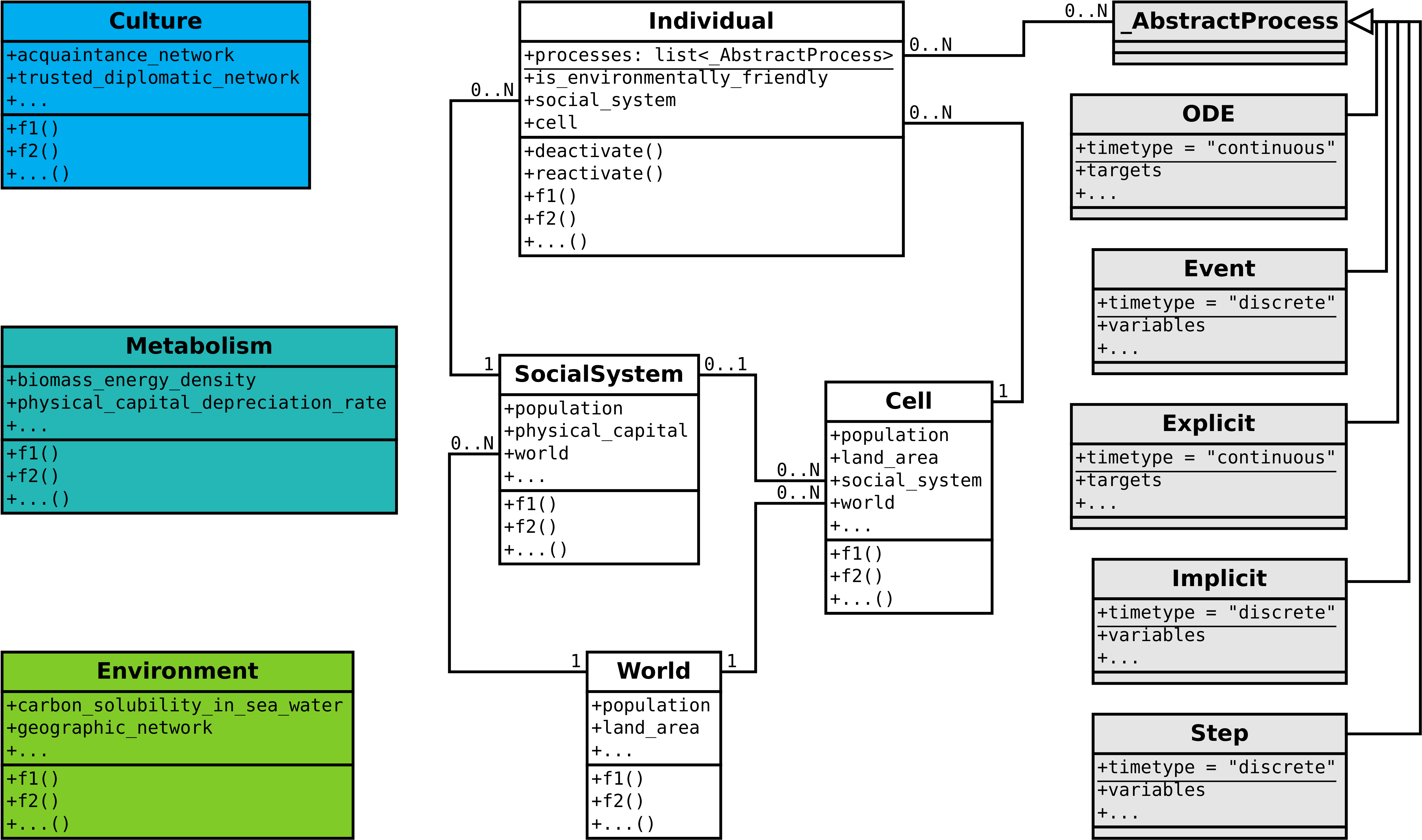}
\caption{\label{fig:basicrelations}%
    \textbf{Basic relationships between entities in the copan:CORE framework.}
    This UML class diagram shows the most important entity types and relationships, 
    and a selection of entities' attributes, 
    as implemented in the `base' model component of the {\em pycopancore} reference implementation. `f1()' and `f2()' are placeholders for process implementation methods belonging to that taxon or entity type. The underlined attributes `processes' (present in all taxa and entity types though shown only once here) and `timetype' are class-level attributes.
}
\end{figure}

We try to keep the number of explicitly considered entity types manageably small
and thus choose to model some relevant things that occur in the real world not as separate entities
but rather as attributes of other entities.
As a rule of thumb (with the exception of the entity type `world'), 
only things that can occur in potentially large, a priori unknown, and maybe changing numbers 
and display a relevant degree of heterogeneity for which a purely statistical description seems inadequate
will be modeled as entities.
In contrast, things that typically occur only once for each entity of some type (e.g., an individual's bank account)
or which are numerous but can sufficiently well described statistically
are modeled as attributes of the latter entity type.

Although further entity types 
(e.g., `household', `firm', `social group', `policy', 
or `river catchment')
will eventually be included into the master data model,
at this point the copan:CORE {\em `base' model component} only provides the entity types which all models must contain, described in this section,
in addition to an overall entity type `world' that may serve as an anchor point for relations between entities
(see also Fig.\,\ref{fig:basicrelations}).

\paragraph{Cells}
An entity of type `cell' represents a small spatial region used for discretising the spatial aspect 
of processes and attributes which are actually continuously distributed in space.
They may be of a more or less regular shape and arrangement, 
e.g., represent a latitude-longitude-regular or an icosahedral grid
or an irregular triangulation adapted to topography. 
Since they have no real-world meaning beyond their use for discretization,
cells are not meant to be used as agents in agent-based model components.
Geographical regions with real-world meaning should instead be modeled via the type `social system'.

\paragraph{Social systems}
An entity of type `social system' is meant to represent what is sometimes simply called a `society', i.e. ``an economic, social, industrial or cultural infrastructure''~\citep{wiki-society} such as a megacity, country, or the EU.
We understand a social system as a human-designed and human-reproduced structure 
including the flows of energy, material, financial and other resources that are used to satisfy human needs and desires, 
influenced by the accessibility and usage of technology and infrastructure~\citep{fischerkowalski1997,otto2018metabolism}.
Equally importantly, social systems include social institutions 
such as informal systems of norms, values and beliefs, 
and formally codified written laws and regulations, governance and organizational structures \citep{Williamson1998}. 
In our framework, norms, values and beliefs may be described in macroscopic terms on the social system level 
but may also be described microscopically on the level of individuals (Sect.\,\ref{sec:entity_individuals}).

Social systems in this sense typically have a considerable size 
(e.g., a sovereign nation state such as the United States of America, a federal state or country such as Scotland, 
an urban area such as the Greater Tokyo Area, 
or an economically very closely integrated world region such as the EU),
controlling a well-defined territory (represented by a set of cells)
and encompassing all the socio-metabolic and cultural processes occurring within that territory.
Social systems are not meant to represent a single social group, class, or stratum,
for which different entity types should be used (e.g., a generic entity type `social group').
To allow for a consistent aggregation of socio-metabolic quantities 
and modeling of hierarchical political decision-making, 
the social systems in a model are either all disjoint 
(e.g., representing twelve world regions as in some integrated assessment models, or all sovereign countries),
or form a nested hierarchy with no nontrivial overlaps 
(e.g., representing a three-level hierarchy of world regions, countries, and urban areas).
As the attributes of social systems will often correspond to data assembled by official statistics,
we encourage to use a set of social systems that is compatible to the standard classification ISO~3166-1/2
when representing real-world social systems.

Social systems may act as agents in agent-based model components
but an alternative choice would be to use `individuals' like their `head of government'  
or `social groups' like a `ruling elite' as agents.

\paragraph{Individuals} \label{sec:entity_individuals}
Entities of type `individual' represent individual human beings.
These entities will typically act as agents in agent-based model components,
although also entities of other types (e.g., the potential types `household', `firm', or 'social group') may do so. 
In contrast to certain economic modeling approaches that use ``representative'' consumers,
an entity of type `individual' in copan:CORE is not usually meant to represent a whole class of similar individuals 
(e.g., all the actual individuals of a certain profession)
but just one specific individual.
Still, the set of all `individuals' contained in a model 
will typically be interpreted as being a representative {\em sample} of all real-world people,
and consequently each individual carries a quantity `represented population' as an attribute
to be used in statistical aggregations, e.g., within a social system.

\subsubsection{Relationships between entity types and process taxa}

Although there is no one-to-one correspondence between process taxa and entity types, some combinations are expected to occur more often than others, as indicated by the thicker gray connections in Fig.\,\ref{fig:si_three-level}.

We expect processes from the {\em environmental (ENV)} process taxon to deal primarily with the entity types 
`cell' (for local processes such as terrestrial vegetation dynamics described with spatial resolution)
and `world' (for global processes described without spatial resolution, e.g., the greenhouse effect)
and sometimes `social system' (for mesoscopic processes described at the level of a social system's territory, 
e.g., the environment diffusion and decomposition of industrial wastes).

{\em Socio-metabolic (MET)} processes are expected to deal primarily with the entity types 
`social system' (e.g., for processes described at national or urban level),
`cell' (for local socio-metabolic processes described with additional spatial resolution for easier coupling to natural processes)
and `world' (for global socio-metabolic processes such as international trade),
and only rarely with the entity type `individual' 
(e.g., for micro-economic model components such as consumption, investment or the job market).

Finally, processes from the {\em socio-cultural (CUL)} taxon are expected to deal primarily with the entity types 
`individual' (for ``micro''-level descriptions) and
`social system' (for ``macro''-level descriptions),
and rarely `world' (for international processes such as diplomacy or treaties).

\subsection{Software design}
\label{sec:core_software}

\begin{figure}[t]
\includegraphics[width=.9 \textwidth]{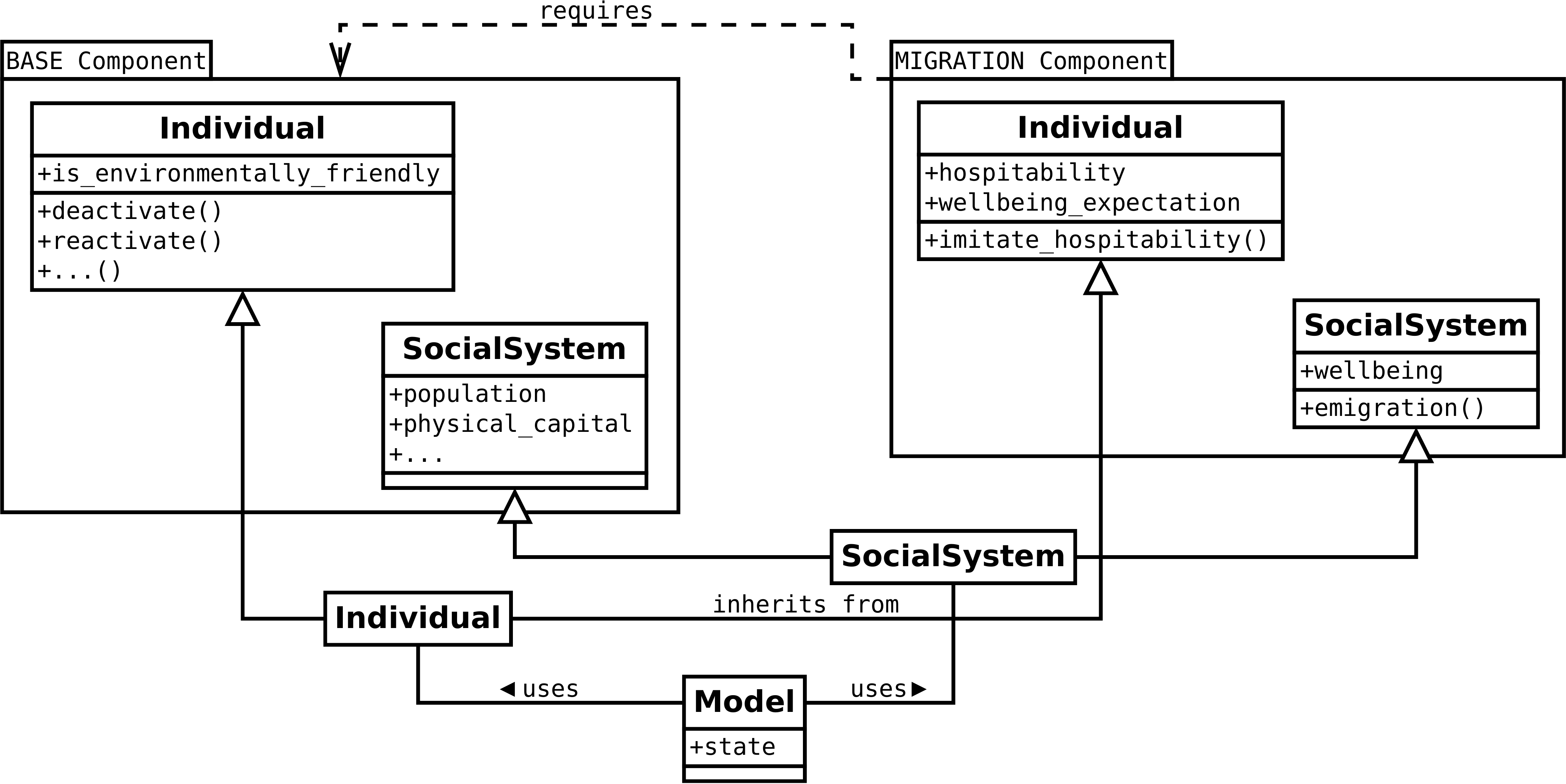}
\caption{%
    \textbf{Model composition through multiple inheritance of attributes and processes by process taxa and entity types.}
    This stylized class diagram shows how a model in copan:CORE can be composed from several model components (only two shown here, the mandatory component `base' and the fictitious component `migration') that contribute component-specific processes and attributes to the model's process taxa and entity types (only two shown here, `Individual' and `SocialSystem'). To achieve this, the classes implementing these entity types on the model level are composed via multiple inheritance (solid arrows) from their component-level counterparts (so-called `mixin' classes).  
    }
\label{fig:si_inheritance}
\end{figure}

This section describes the programming language-independent parts of 
how the above abstract structure is realized as computer software.
As they correspond closely with the role-based and entity-centric view of the abstract framework,
{\em modularization} and {\em object-orientation} are our main design principles.
All parts of the software are organized in packages, subpackages, modules, and classes.
The only exception are those parts of the software that are written by model end-users to perform actual studies, 
which will typically be in the form of {\em scripts} 
following a mainly imperative programming style that uses the classes provided by the framework.
Fig.\,\ref{fig:si_inheritance} summarizes the main aspects of this design which are described in detail in the following.

\paragraph{Object-oriented representation}
Entity types and process taxa are represented by {\em classes} (`Cell', `SocialSystem', `Culture', \ldots),
individual entities by {\em instances} (objects) of the respective entity type class,
and process taxon classes have exactly one instance.
While entity type and process taxon classes hold processes' and variables' metadata as {\em class attributes,} 
entity instances hold variable values and, where needed, their time derivatives as {\em instance attributes.}
Processes' logics can be specified via {\em symbolic expressions} in the process metadata 
(e.g., for simple algebraic or differential equations)
or as imperative code in {\em instance methods}
(e.g., for regular `steps' and random `events' in an agent-based modeling style),
thereby providing a large flexibility in how the equations and rules of the model are actually represented in the code, without compromising the interoperability of model components.

\paragraph{Interface and implementation classes}
All of this is true not only on the level of (composed) models
but already on the level of model components, 
though restricted to the entity types, processes and variables used in the respective component.
To avoid name clashes but still be able to use the same simple naming convention throughout in all model components,
each model component is represented by a {\em subpackage} of the main copan:CORE software package, 
containing class definitions for all used entity types and process taxa as follows.
Each entity type and process taxon used in the model component is represented by two classes,
(i) an {\em interface class} that has a class attribute of type `Variable' 
(often imported from the master data model subpackage or another model component's interface classes)
for each variable of this entity type or process taxon this model component uses as input or output, 
containing that variable's metadata (see Fig.\,1 in the Supplementary Information for an example),
and (ii) an {\em implementation class} inherited from the interface class,
containing a class attribute `processes' and potentially some instance methods with process logics.

The attribute `processes' is a list of objects of type `Process', 
each of which specifies the metadata of one process that this model component contributes to this entity type or process taxon
(see Figs. 
2 and 3 of the Supplementary Information
for examples).
These metadata either contain the process logics as a symbolic expression or as a reference to some instance method(s).
Instance methods do not return variable values but manipulate variable values or time derivatives directly 
via the respective instance attributes.
As many variables are influenced by more than one process,
some process implementation methods (e.g., those for differential equations or noise) 
only add some amount to an attribute value, 
while others (e.g., those for major events) may also overwrite an attribute value completely.

\paragraph{Model composition via multiple inheritance}
Finally, a model's composition from model components is represented via multiple {\em inheritance} 
from the model component's implementation classes (which are thus also called `mixin' classes) as follows.
Each model is defined in a separate {\em module} (typically a single code file).
For each entity type and process taxon that is defined in at least one of the used model component packages,
the model module defines a composite class that inherits 
from all the mixin classes of that entity type contained in the used model component packages.
Fig.\,\ref{fig:si_inheritance} shows an example of this with just two components and two entity types.

\paragraph{Dimensional quantities, symbolic expressions, networks}
To be able to specify values of dimensional quantities, mathematical equations,
and networks of relationships between entities in a convenient and transparent way,
we provide classes representing these types of objects,
e.g., `Dimension', `Unit', `DimensionalQuantity', `Expr' (for symbolic expressions), `Graph' (for networks),
`ReferenceVariable'/`SetVariable' (for references to single/sets of other entities).

\paragraph{Interoperability with other model software}
copan:CORE can be used together with other simulation software to simulate coupled models consisting of ``internal'' components implemented in copan:CORE interacting in both directions with an ``external'' component provided by the other software.
Currently, copan:CORE must act as the coupler to achieve this, which requires that the other software provides at least a minimal interface (e.g., conforming to the basic modeling interface (BMI), \citet{Syvitski2014}) that allows to read, set and change its state variables and to advance its model simulation by one time step.

To couple an external model component into a copan:CORE model, one must write a ``wrapper'' model component in the copan:CORE framework. 
For each relevant `external' variable of the external model, the wrapper specifies a corresponding `internal' copan:CORE variable in a suitable entity type or process taxon. 
In addition, the wrapper contributes a process implementation method of type `Step' to a suitable process taxon, which uses the external software's interface to sync the external variables with their internal versions, using a suitable regridding strategy if necessary, and lets the external model perform a time step.

In later versions, copan:CORE will include a standard wrapper template for models providing a BMI, and might also itself provide such an interface to external couplers.

\subsection{Reference implementation in Python}
\label{sec:core_python}

For the reference implementation of copan:CORE we chose the Python programming language 
to enable a fast development cycle and provide a low threshold for end users. 
It is available as the open-source Python package {\em pycopancore} (\texttt{\url{https://github.com/pik-copan/pycopancore}})
including the master data model 
and a small number of pre-defined model components and models as subpackages and modules.
Symbolic expressions are implemented via the {\em sympy} package~\citep{Sympy2017} 
which was extended to support aggregation (as in Fig.\,
3 of the Supplementary Information,
top, line 5) 
and cross-referencing between entities 
(same Fig., bottom, line 14).
ODE integration is currently implemented via the {\em scipy} package~\citep{Jones2001scipy}.
While the reference implementation is suitable for moderately sized projects,
very detailed models or large-scale Monte-Carlo simulations 
may require an implementation in a faster language such as C++, 
which we aim at realizing via a community-driven open-source software development project.
Fig.\,\ref{fig:script} gives an impression of how user code in \textit{pycopancore} looks like. See the \emph{Supplementary Information} for further details.

\begin{figure}
\centering
\includegraphics[width=0.9\textwidth,clip,trim=1in 5.2in 0in 1.8in]{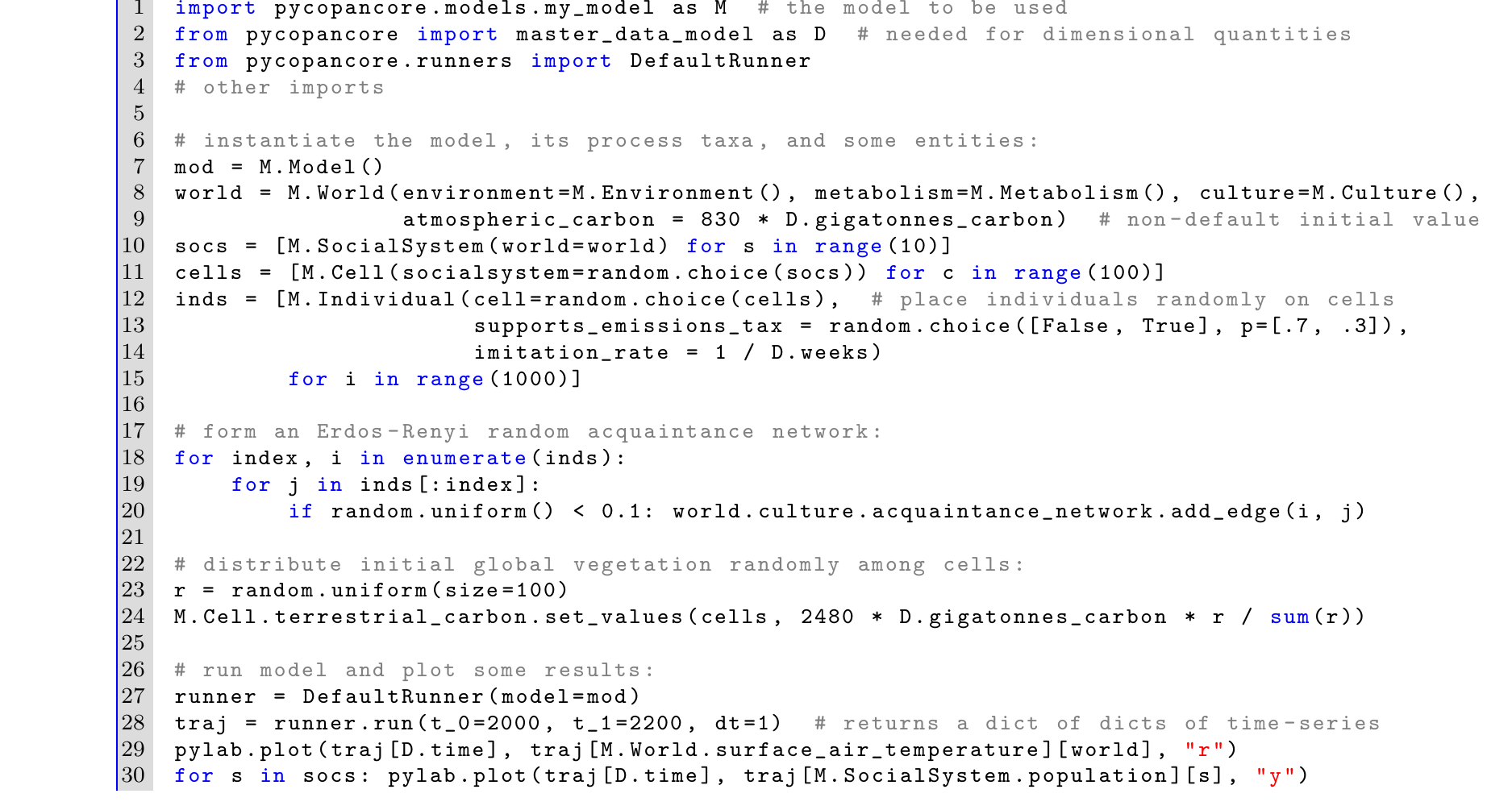}
\caption{\label{fig:script}
\textbf{Sketch of a model end user's Python script running a model and plotting some results},
featuring dimensional quantities and a network. 
Variable values can be set either at instantiation (line 9),
via the entity object attribute (line 20) or the Variable object (line 24).
}
\end{figure}

\begin{figure} 
\small\flushleft
\centering
\includegraphics[width=0.9\textwidth,clip,trim=1in 5.5in 0.5in 1.7in]{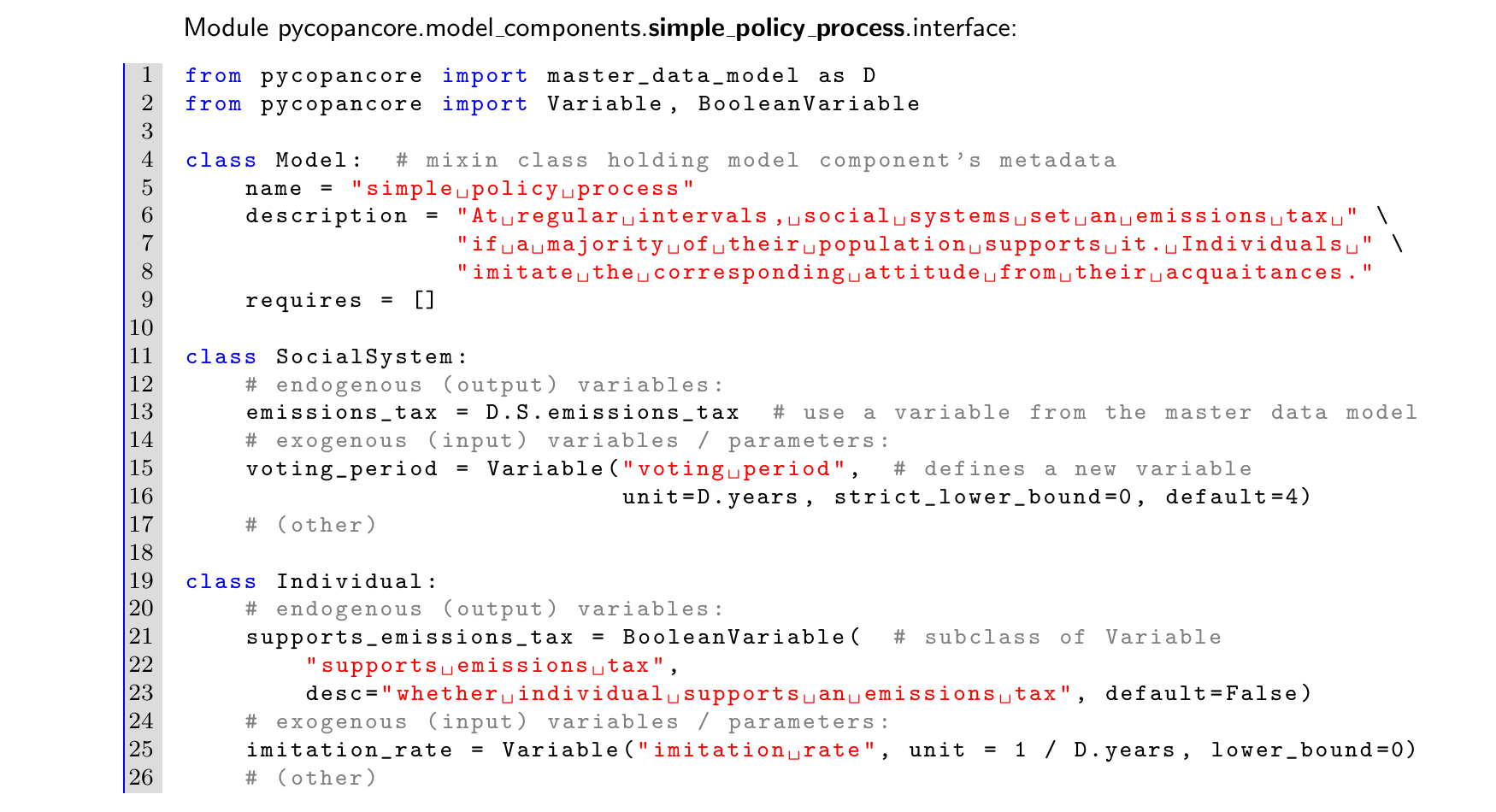}
\caption{\label{SIfig:interface}
    Sketch of a model component's interface, implemented as a Python module that lists the variables. 
    The component contributes to the various entity types and process taxa,
    either referenced from the master data model (line 13) or defined newly (line 15).
}
\end{figure}

\begin{figure} 
\centering
\includegraphics[width=0.9\textwidth,clip,trim=1in 2in 0.5in 1.7in]{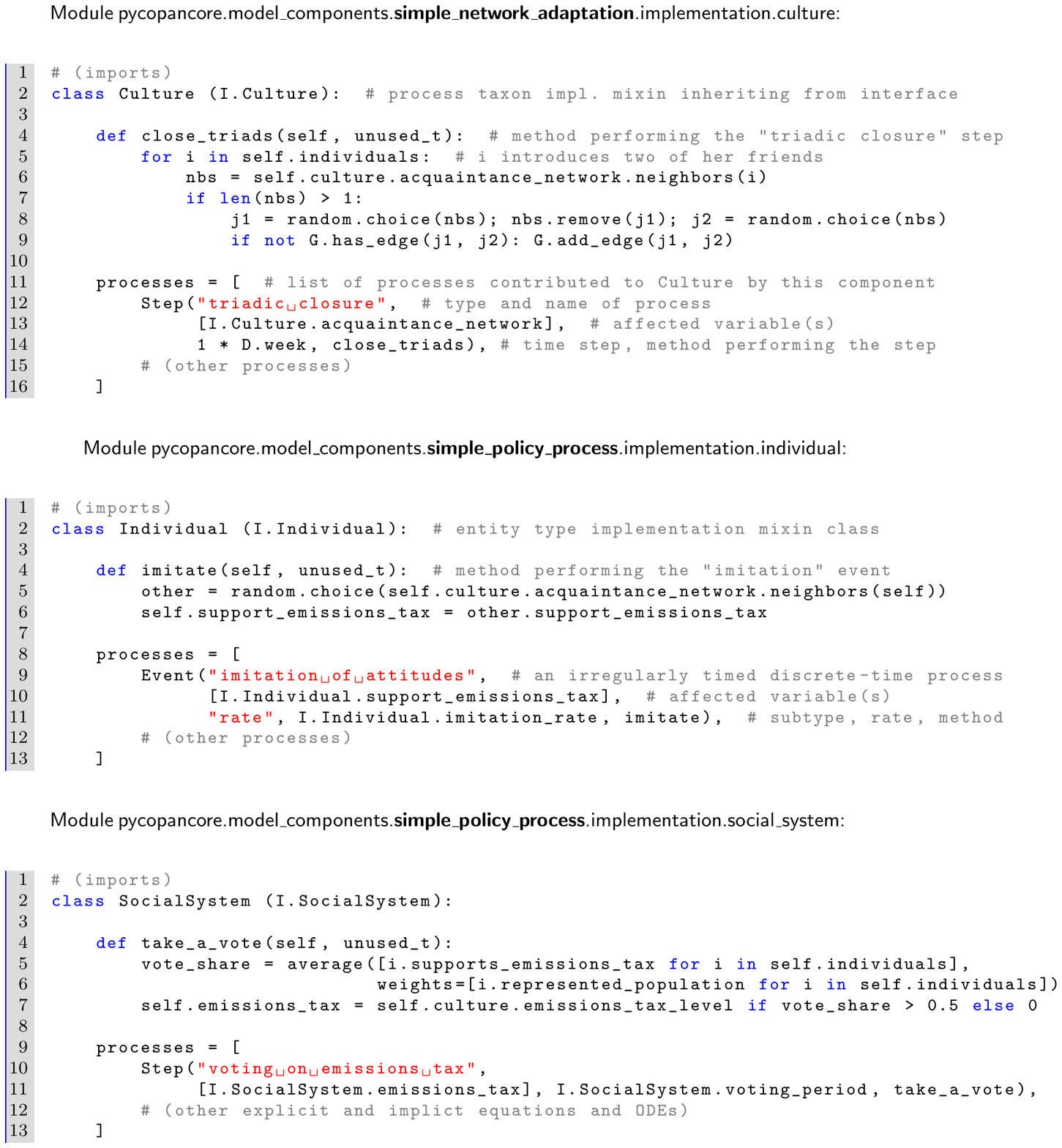}
\caption{\label{SIfig:implementation1}
    Sketches of implementation classes for three entity-types in two    model components, 
    to be used as mixin classes in model composition. 
    Each class defines processes (here steps and events) 
    that the owning model component contributes to a certain process taxon or entity type.
    Note how the examples feature
    process implementation via instance methods (l.4 of each example) 
    networks (top, l.6--9), dimensional quantities (top, l.14),
    stochasticity (middle, l.5), 
    and the use of a social system's individuals as a representative sample of its population (bottom, l.5+6).
    See inline comments in magenta for detailed explanations.
}
\end{figure}

\begin{figure} 
\centering
\includegraphics[width=0.9\textwidth,clip,trim=1in 4.2in 0.5in 1.7in]{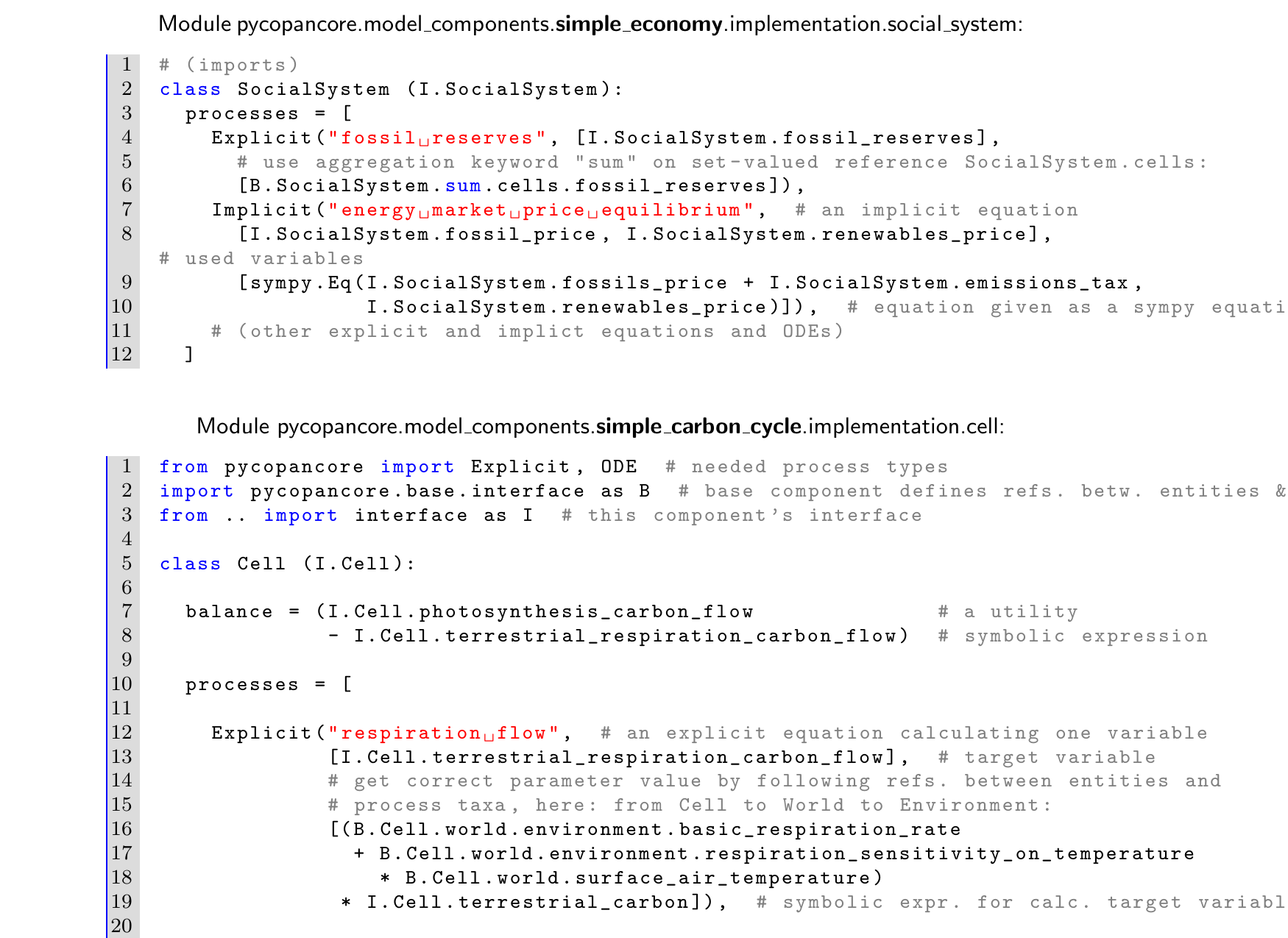}
\caption{\label{SIfig:implementation2}
    Sketches of implementation classes (continued),
    featuring explicit and implicit equations (top, l.4--10) and ODEs (bottom, l.23--26), 
    symbolic expressions (bottom, l.7--8) and equations (top, l.9--10), 
    aggregation (top, l.6), 
    and cross-referencing between entities (bottom, l.16--18).
}
\end{figure}

{\bf Model component developers} 
add their model component as a package to their local workspace folder, 
including the interface module and one module for each provided implementation class
as in Figs.\,\ref{SIfig:interface}--\ref{SIfig:implementation2}. 
In order to paralellize complex computations or couple to other model software, 
an implementation class may implement certain processes using an instance method 
that calls or communicates with other processes or external programs 
which provide a communication method supported by Python, e.g., MPI or JSON.

{\bf Model composers}
provide a module that mainly composes the final entity type and process taxon classes 
via multiple inheritance from model components' implementation classes,
e.g., specifying code like
{\small
\begin{verbatim}
    import .climate_policy as pol
    import pycopancore.model_components.simple_economy as econ 
    import pycopancore.base
    class SocialSystem (pol.SocialSystem, econ.SocialSystem, base.SocialSystem): 
        pass
\end{verbatim}
}
{\bf Model end users}
use a Python script that imports these model modules,
instantiates a `model' object, all needed process taxon objects and an initial set of entities, 
then initializes those variables that shall start with non-default values,
uses a `runner' object to run the model for a specified time
and finally analyses the resulting trajectory.
Fig.\,5 of the main text gives a sketch of such a script
(see the online tutorial for more detailed examples).

Upon instantiation, the `model' object uses Python's {\em introspection} capabilities
to analyse its own model structure including which variables depend on which others in which way,
and this information is then used by the runner to simulate the model.
Future versions will use this information further for improving performance and producing reports on model structure.
The runner returns the time evolution of requested variables as a nested Python dictionary 
the first- and second-level keys of which are a `variable' object and an entity or process taxon 
and whose values are lists of values ordered by time, 
which can then conveniently be analysed or plotted 
(e.g., Fig.\,5 of the main text, line 30). 

\section{Details and potential extensions of the example model}

For variables and parameters taken from \cite{Nitzbon2017} 
we chose values equivalent to those given in Table~1 of that paper,
mainly except for
$l_0$, which was chosen higher to accommodate for our additional space competition factor and make the initial photosynthesis flow fit current amounts. As initial conditions, we used rough global aggregates where data was available.

The model presented in the main text can easily be extended to include well-being-driven population growth and migration, renewable technology knowledge spillovers, and carbon taxation, for all of which the reference implementation shippes with corresponding model components.

\subsection{Population growth}

Like in \cite{Nitzbon2017}, but here again on the {\tt SocialSystem} level,
population has a wellbeing-dependent fertility rate that was roughly fitted against
country-level data of fertility vs GDP per capita.
We chose the functional form
${\rm fert}_s = p_0 + 2 (p - p_0) W_s W_P^{\omega_p} / (W_s^{1+{\omega_p}} + W_P^{1+{\omega_p}})$,
where wellbeing $W_s = w_Y (1 - i) Y_s / P_s + w_L L_s / \Sigma_s$ 
depends on per-capita consumption $(1 - i) Y_s / P_s$ 
and the mean terrestrial carbon density in that social system, $L_s / \Sigma_s$.
For small $W_s$, $f$ grows linearly, reaching its maximum at $W_s = W_P$,
then decaying towards $p_0$ with an asymptotically power-law shape 
with exponent $\omega_p$ for large $W_s$.
Similarly, for mortality we roughly fitted the function
$q / (W_s / W_P)^{\omega_q}$ against data
and added a term representing increased mortality from climate change impacts,
$q_T (T - T_q)$,
and one representing competition for space,
$q_C P_s / \Sigma_s \sqrt{K_s}$,
where the factor $1 / \sqrt{K_s}$ represents the assumption that
housing is a form of physical capital with decreasing marginal value.

Note that while population $P_s$ changes over time,
the number of representative individuals in $s$ remains constant in our example model,
implying that the share of the population in $s$ 
that a certain {\tt Individual} $i$ represents will change over time.
A more elaborate model could try to keep the ratio of population 
and number of representative individuals roughly constant
by generating or deactivating instances of {\tt Individual} in $s$ at the current
birth and death rate of $s$.

\subsection{Wellbeing-driven migration}

In addition to births and deaths, 
{\tt SocialSystem}s' populations change due to migration depending on differences in wellbeing.
We assume each person in $s$ has a probability of emigrating to $s'$ that is proportional to 
the available information about differences in wellbeing for which we use the population in $s'$' as a proxy. 
We also assume that the probability of migration depends on a sigmoidal function of the wellbeing ratio, 
$f(\log W_{s'} - \log W_s)$ with $f(-\infty) = 0$ and $f(\infty) = 1$.
More specifically, the absolute emigration flow from $s$ to $s'$ is
$\mu P_s P_{s'} (1/2 + \arctan (\pi \phi (\log W_{s'} - \log W_s - \log \rho)) / \pi)$,
where $\mu$ is a basic rate and $\phi$ and $\rho$ are slope and offset parameters.

\subsection{Carbon taxation}

In the general elections of a {\tt SocialSystem}, also a GHG emissions tax may be introduced, leading to a shift in the energy price equilibrium of 
\begin{align*}
    &\text{marginal production cost of fossil energy} + \text{emissions tax} \\
    &= \text{marginal production cost of biomass energy} + \text{emissions tax} \\
    &= \text{marginal production cost of renewable energy} - \text{renewable subsidy}.
\end{align*}

\begin{sidewaysfigure}
\centering
\includegraphics[width=\textwidth]{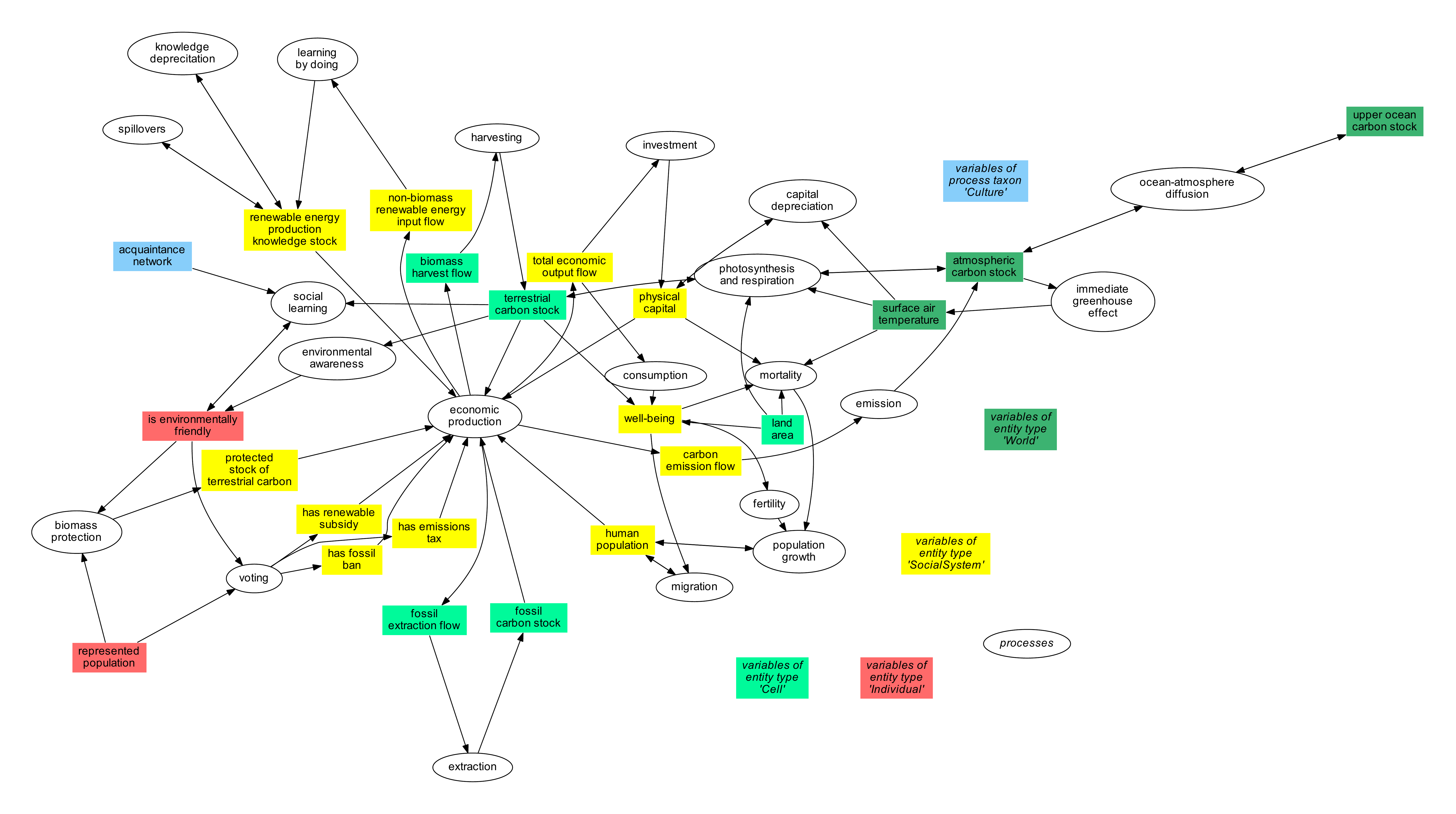}
\caption{\label{fig:vars_procs}
    Main variables and processes of the example model and its extension.
}
\end{sidewaysfigure}

\subsection{Necessary improvements}

We'd like to repeat that 
the example model was designed to showcase the concepts and capabilities of copan:CORE in a rather simple WEM, 
and its components were chosen so that all entity types and process taxa and most features of copan:CORE are covered.
The example model is not intended to be a serious representation of the real world 
that could be used directly for studying research questions,
and the shown time evolutions may not be interpreted as any kind of meaningful quantitative prediction or projection.

To develop the example model into a serious World-Earth model, 
very many things remain to be done,
including a careful selection of processes to include or exclude,
improvements in model equations and agent's behavioural rules,
both by fitting data where available 
and adopting model components from the literature, 
suitable choice of real-world social systems to include as entities,
appropriate gridding of the surface into cells,
and a solid estimation of parameters and initial conditions 
and their local and societal differences.

\bibliographystyle{copernicus} 
\bibliography{copancore_si.bib,copancore_literature.bib,literature_jobst.bib,literature_jona.bib}